\documentclass[preprint,12pt,authoryear]{elsarticle}

\usepackage[british]{babel}
\usepackage{tikz}
\usepackage{pgfplots}
\pgfplotsset{compat=newest}
\usepgfplotslibrary{groupplots}
\usepgfplotslibrary{dateplot}
\usepackage{sansmath}
\usepackage{enumitem}
\usepackage[ps2pdf,bookmarksopen=true,citebordercolor={1 1 1},urlbordercolor={1 1 1},linkbordercolor={1 1 1}]{hyperref}
\hypersetup{hidelinks,breaklinks}

\usepackage[linesnumbered,ruled,vlined]{algorithm2e}

\SetKwInput{KwInput}{Input}                
\SetKwInput{KwOutput}{Output}              

\usepackage{amssymb}
\usepackage{amsmath}

\usepackage{amsthm}

\usepackage{acronym}
\acrodef{OFO}{Online Feedback Optimization}

\usepackage{dsfont}
\usepackage{bm}
\usepackage{booktabs}
\usepackage[width=.45\textwidth]{caption}
\usepackage{subcaption}
\usepackage{color}
\usepackage{epstopdf}
\usepackage{cleveref}

\usepackage{booktabs}
\usepackage{multirow}
\usepackage[normalem]{ulem}
\useunder{\uline}{\ul}{}
\usepackage{afterpage} 
\usepackage{longtable}
\usepackage{psfrag}
\usepackage{array}
\usepackage{textcomp}

\usepackage{floatrow}
\DeclareFloatFont{tiny}{\tiny}
\restylefloat{figure}

\usepackage{float}
\floatstyle{plaintop}
\restylefloat{table}

\usepackage{graphicx}
\graphicspath{{Figures/}}

\newcommand{\T}{^\mathsf{T}} 
\renewcommand{\d}{\mathrm{d}} 
\newcommand{\R}{\mathds{R}} 

\DeclareMathOperator*{\argmin}{arg\,min}
\DeclareMathOperator*{\argmax}{arg\,max}

\journal{Journal of Process Control}

\usepackage{tikz}
\usetikzlibrary{plotmarks}
\usetikzlibrary{shapes,snakes}
\usepackage{pgfplots}

\begin{document}

\begin{frontmatter}

\title{Online Feedback Optimization of Compressor Stations\\with Model Adaptation using\\Gaussian Process Regression}

\author[ICL,ETH]{M. Zagorowska}
\author[ETH]{M. Degner}
\author[ETH]{L. Ortmann}
\author[ICL]{A. Ahmed}
\author[ETH]{S. Bolognani}
\author[ICL]{E. A. del Rio Chanona}
\author[ICL]{M. Mercang{\"o}z}

\address[ICL]{Department of~Chemical Engineering,	Imperial College London, South Kensington, SW7 2AZ London, UK (m.zagorowska@imperial.ac.uk, a.ahmed21@imperial.ac.uk, a.del-rio-chanona@imperial.ac.uk, m.mercangoz@imperial.ac.uk)}
\address[ETH]{Automatic Control Laboratory, ETH Zurich, 8092 Zurich, Switzerland (mzagorowska@control.ee.ethz.ch, mdegner@student.ethz.ch, ortmannl@control.ee.ethz.ch, bsaverio@ethz.ch)}

\begin{abstract}
Online Feedback Optimization is a method used to steer the operation of a process plant to its optimal operating point without explicitly solving a nonlinear constrained optimization problem. This is achieved by leveraging a linear plant model and feedback from measurements. However the presence of plant-model mismatch leads to suboptimal results when using this approach. Learning the plant-model mismatch enables Online Feedback Optimization to overcome this shortcoming. In this work we present a novel application of Online Feedback Optimization with online model adaptation using Gaussian Process regression. We demonstrate our approach with a realistic load sharing problem in a compressor station with parametric and structural plant-model mismatch. We assume imperfect knowledge of the compressor maps and design an Online Feedback Optimization controller that minimizes the compressor station power consumption. In the evaluated scenario, imperfect knowledge of the plant leads to a 5\% increase in power consumption compared to the case with perfect knowledge. We demonstrate that Online Feedback Optimization with model adaptation reduces this increase to only 0.8\%, closely approximating the case of perfect knowledge of the plant, regardless of the type of mismatch.

\end{abstract}


\end{frontmatter}

\section{Introduction}
A model provides a mathematical description of how an industrial system reacts to its surroundings, changes in its state, and its control inputs. Due to the complexity of physical processes governing the behaviour of industrial systems, mathematical models differ from real systems. This difference between a mathematical model and a real system is called a \emph{plant-model mismatch}. Model-based optimization solutions for the operation of industrial systems must take this mismatch into account. In this paper we provide a method for online mitigation of plant-model mismatch combining Gaussian process regression with Online Feedback Optimization. The results are shown for load sharing optimization in a realistic compressor station with centrifugal compressors working in parallel.

\subsection{Industrial motivation}
There is renewed interest to increase the energy-efficiency of industrial operations as a quick way for reducing the carbon emissions associated with the consumed energy \citep{Riungu_2022,Lei_2021}. Replacing the motors of rotating machines with more efficient alternatives and converting them to variable speed operation is an ongoing trend. The control and optimization of these machines will be a key technology to capture the full energy saving benefits coming from the increased range of speeds and loads provided by variable speed operation. That said, such machines are typically controlled by embedded units such as programmable logic controllers or so called edge computing devices, which tend to have relatively limited computational capabilities. In this regard, \ac{OFO} is a model-based method that can steer such a machine or a collection of these machines to an optimum point without solving a demanding nonlinear constrained optimization problem \citep{Optimization_Hauswirth2021}. 

The performance of rotating machinery tends to degrade over time and accurate models for such systems are generally not available \citep{Influence_Zagorowska2019}. Large project efforts to develop such accurate models could discourage operators -- especially small and medium size ones -- and prevent the reduction of their carbon footprint. In this paper, we describe how we can combine online learning via Gaussian Process regression with \ac{OFO} to enable the steering of a plant to the optimal operating point with an inaccurate model for \ac{OFO} calculations. The described approach can be deployed at the edge with moderate modelling effort.

\subsection{Background}

\subsubsection{Compressor stations}
A compressor station is typically a part of a natural gas transport network and provides a boost for transporting the gas to the receivers \citep{BS_BSI2014}. The objective for the operation of a compressor station is to satisfy varying demand by adjusting how much gas is processed by each compressor. The amount of gas processed by a compressor is called a \textit{load}. The process of assigning the loads to the compressors is called \emph{load sharing}. The primary objective of load sharing is to allocate the loads to compressors in a compressor station to minimise the operating cost. The behaviour of a centrifugal compressor is usually modelled by its characteristics capturing how the pressure, compressor load, and compressor speed are connected. Due to the inherent differences between the characteristics of the compressors, the equal load approach is not optimal  \citep[Ch. 8.15]{Instrument_Liptak2005}, \citep{Influence_Zagorowska2019}. The loads should be adjusted to the nonlinear characteristics of the compressors so that the compressors stay within their operating ranges and that the demand is satisfied.

\subsubsection{Plant-model mismatch}

The load-sharing problem with varying compressor characteristics was analysed by \cite{Load_Paparella2013}, \cite{Optimal_Milosavljevic2016}, and \cite{Load_Kumar2017}. They all approximated the characteristics of a compressor using polynomial functions and updated the parameters online to match the approximation to the real system. The updated characteristics of compressor efficiency and head were then used for optimal load-sharing. In particular, \cite{Optimal_Xenos2015} used piecewise linear approximation of compressor maps to capture the inherent nonlinear nature of their operation. The performance of the optimization and, in consequence, of the compressor station depends on how well the models capture the real characteristics. The approaches used in the literature have limited capabilities for compensating structural mismatch if the real characteristics has a different functional form than the assumed model. The current work uses Gaussian processes to estimate the difference between the real characteristics and the model, which enables good performance in the case of parametric as well as structural mismatch.

Nonlinear characteristics of compressors are used by multiple authors over the years, including \cite{Local_Osiadacz1981}, \cite{Optimized_Jenicek1995}, \cite{Model_Wu2000} and recently \cite{Improved_Jung2017} who take into account individual characteristics of each compressor when solving the load sharing optimization problem. However, they assume perfect knowledge about the characteristics. Taking into account that the behaviour of a compressor might be different than expected, \cite{Online_Cortinovis2016} and \cite{Real_Milosavljevic2020} have designed model-based optimizing control for a compressor station. They assume that the characteristics of a compressor change over time. For optimization, they approximate the characteristics with a polynomial and identify the parameters of the polynomial online. However, their approach relies on the degree of the polynomial used for identification and may result in low model fidelity, and in consequence, suboptimal performance. The assumption about the fixed polynomial degree was relaxed by \cite{Real_Gentsch2020} who formulated real-time approximation of compressor characteristics as a dynamic optimization problem. They have used a Kalman filter to estimate online the properties of the gas and the dynamics of the compressors. The estimated quantities can then be used to compute the characteristics from first principles. Their approach is independent from the functional form of the characteristics, but requires multiple measurements for improved accuracy of the estimated characteristics. In the current work we mitigate the need for multiple measurements by using Gaussian process regression.

\subsubsection{Gaussian process regression}

Gaussian processes regression is a non-parametric generalisation of standard linear regression \citep{Gaussian_Rasmussen2006}. The flexibility of Gaussian process and their non-parametric nature make them useful for applications where the underlying functional form of the model is unknown. Gaussian processes have been used for modelling compressor characteristics by \cite{Improved_Chu2018} and \cite{Centrifugal_Vilalta2019} who indicate that Gaussian process regression enables quick learning of the unknown model of compressor characteristics. The work by \cite{Application_Ahmed2021} further demonstrates that Gaussian process regression captures well the characteristics of a compressor even if only few data points are available. As a result, the performance of Gaussian process regression is better than linear regression or polynomial fitting used by \cite{Optimal_Milosavljevic2016}. In the current work, we explore the Gaussian process regression for compressors from \cite{Application_Ahmed2021} in a novel control method called Online Feedback Optimization that uses online measurements to drive the system to optimal operating points. 

\subsubsection{Online Feedback Optimization}

\cite{Optimization_Hauswirth2021} have presented an overview of feedback controllers that use measured data to mitigate the dependence of the controller on model fidelity. They describe an approach called Online Feedback Optimization, which merges feedback control with gradient-based optimization algorithms. In contrast to existing methods for optimizing operation of complex processes, such as real-time optimization or modifier adaptation, Online Feedback Optimization exploits the properties of feedback control and iterative optimization algorithms to ensure reaching the optimum. In particular, if the dynamics of the optimized system are fast, existing methods are inefficient because they require solving complex nonlinear optimization at every timestep. By relying only on measured outputs, feedback optimization provides computationally efficient solutions to complex problems with fast dynamics. As the dynamics of the compressors are faster than the changes in demand, OFO is a good candidate for designing operation of compressor stations. \cite{Online_Degner2021} presents a possible application of \ac{OFO} to load-sharing in compressor stations, assuming that the gradients are available and there is no mismatch between the plant and the model. In the current work, we merge OFO with Gaussian process regression to compensate for the mismatch in a compressor station. By using the fact that OFO enables avoiding explicit solution of nonlinear optimization problems we solve the load-sharing problem as a fully online problem.

\subsection{Contributions}

The novelty of the current paper consists in explicit improvement of the model of compressors used in \ac{OFO} by performing online Gaussian process regression to mitigate plant-model mismatch. The contributions of the current paper are:
\begin{itemize}
    \item Formulation of a load-sharing problem in a compressor station with unknown characteristics as OFO, reducing the need for detailed knowledge about the nonlinear model of the compressors;
    \item Demonstration of online Gaussian process regression as a way of estimating performance characteristics for gradient-based optimization;
    \item Application of OFO with online Gaussian process regression in a simulation of a realistic compressor station with parametric and structural mismatch.
\end{itemize}

This paper is structured as follows. Section \ref{sec:CompressorStations} describes the compressor station used in this paper. Section \ref{sec:GPefficiency} presents the model adaptation using Gaussian process regression, which is then used in Section \ref{sec:FOcompressors} in \ac{OFO}. Section \ref{sec:Results} presents the results of \ac{OFO} with model adaptation in the compressor station. The results and potential future research areas are discussed in Section \ref{sec:Limitations}. The paper ends with conclusions in Section \ref{sec:Conclusions}.

\section{Compressor station}
\label{sec:CompressorStations}

\begin{figure}
\centering
\begin{tikzpicture}[scale=0.8]
\draw [ultra thick,dashed, -{stealth[scale=2]}] (-1,4) -- (0,4);
\node at (-1,4.5) {\scriptsize{\textsf{Demand}}};
\draw (0,3.5) rectangle (1,4.5);
\node at (0.5,4) {\scriptsize{\textsf{LS}}};
\draw [ultra thick,dashed, -{stealth[scale=2]}] (1,4) -- (1.9,4);

\draw[fill] (6,10.5) arc [radius = 0.2, start angle=0, end angle= 180];
\draw [ultra thick] (5.8,10.5) -- (5.8,10.35);
\draw [ultra thick] (5.6,10.4) -- (6,10.2);
\draw [ultra thick] (6,10.4) -- (5.6,10.2);
\draw [ultra thick] (5.6,10.4) -- (5.6,10.2);
\draw [ultra thick] (6,10.4) -- (6,10.2);
\draw [ultra thick] (6.5,9) -- (6.5,10.3);
\draw [ultra thick] (6.5,10.3) -- (6,10.3);
\draw [ultra thick] (5.6,10.3) -- (4.5,10.3);
\draw [ultra thick] (4.5,10.3) -- (4.5,9.8);
\draw [ultra thick, -{stealth[scale=2]}] (4.5,9.8) -- (4.9,9.8);

\draw[fill] (6,6.5) arc [radius = 0.2, start angle=0, end angle= 180];
\draw [ultra thick] (5.8,6.5) -- (5.8,6.35);
\draw [ultra thick] (5.6,6.4) -- (6,6.2);
\draw [ultra thick] (6,6.4) -- (5.6,6.2);
\draw [ultra thick] (5.6,6.4) -- (5.6,6.2);
\draw [ultra thick] (6,6.4) -- (6,6.2);
\draw [ultra thick] (5.6,6.4) -- (5.6,6.2);
\draw [ultra thick] (6,6.4) -- (6,6.2);
\draw [ultra thick] (6.5,5) -- (6.5,6.3);
\draw [ultra thick] (6.5,6.3) -- (6,6.3);
\draw [ultra thick] (5.6,6.3) -- (4.5,6.3);
\draw [ultra thick] (4.5,6.3) -- (4.5,5.8);
\draw [ultra thick, -{stealth[scale=2]}] (4.5,5.8) -- (4.9,5.8);
\node [below] at (7,9) {$m_1^c$};

\draw[fill] (6,2.5) arc [radius = 0.2, start angle=0, end angle= 180];
\draw [ultra thick] (5.8,2.5) -- (5.8,2.35);
\draw [ultra thick] (5.6,2.4) -- (6,2.2);
\draw [ultra thick] (6,2.4) -- (5.6,2.2);
\draw [ultra thick] (5.6,2.4) -- (5.6,2.2);
\draw [ultra thick] (6,2.4) -- (6,2.2);
\draw [ultra thick] (6.5,1) -- (6.5,2.3);
\draw [ultra thick] (6.5,2.3) -- (6,2.3);
\draw [ultra thick] (5.6,2.3) -- (4.5,2.3);
\draw [ultra thick] (4.5,2.3) -- (4.5,1.8);
\draw [ultra thick, -{stealth[scale=2]}] (4.5,1.8) -- (4.9,1.8);
\node [below] at (7,5) {$m_2^c$};

\draw[fill] (2,4) circle [radius=0.1];
\draw [dashed,ultra thick] (2,5) -- (2,-1);
\draw [ultra thick] (6,9) -- (9,9);
\draw [ultra thick] (6,1) -- (9,1);
\draw [ultra thick] (6,5) -- (9,5);
\draw [ultra thick, -{stealth[scale=2]}] (1.5,9.25) -- (5,9.25);
\draw [ultra thick, -{stealth[scale=2]}] (1.5,5.25) -- (5,5.25);
\draw [ultra thick, -{stealth[scale=2]}] (1.5,1.25) -- (5,1.25);
\draw [dashed,ultra thick] (2,5) -- (2,7);
\draw [dashed,ultra thick, -{stealth[scale=2]}] (2,-1) -- (3.2,-1);
\draw [dashed,ultra thick, -{stealth[scale=2]}] (2,7) -- (3.2,7);
\draw [dashed,ultra thick, -{stealth[scale=2]}] (2,3) -- (3.2,3);
\draw [dashed,ultra thick,-{stealth[scale=2]}] (3.55,7.5) -- (3.55,8.2);
\draw [dashed,ultra thick,-{stealth[scale=2]}] (3.55,3.5) -- (3.55,4.2);
\draw [dashed,ultra thick,-{stealth[scale=2]}] (3.55,-0.5) -- (3.55,0.2);
\draw [dashed,ultra thick,-{stealth[scale=2]}] (4,8.55) -- (5,8.55);
\draw [dashed,ultra thick,-{stealth[scale=2]}] (4,4.55) -- (5,4.55);
\draw [dashed,ultra thick,-{stealth[scale=2]}] (4,0.55) -- (5,0.55);
\node [below] at (7,1) {$m_3^c$};

\node[ultra thick,
    trapezium, 
    draw,
    rotate=-90,
    trapezium stretches = true,
    minimum width = 1.2cm, 
    minimum height = 0.7cm
    ]
    at (5.5,9) {};

\node at (5.5,9) {\scriptsize{\textsf{C1}}};

\draw (3.2,7.5) rectangle (3.9,6.8);
\node at (3.55,7.1) {\scriptsize{\textsf{FC}}};

\draw [rounded corners] (3.0,8.9) rectangle (4.0,8.2);
\node at (3.5,8.6) {\scriptsize{\textsf{D}}};
\node [right] at (3.55,7.8) {$\omega_{1}$};
\node [above] at (4.5,8.5) {$\tau_{1}$};

\draw [dashed] (7,11.5) rectangle (8.1,10.8);
\node at (7.5,11.15) {\scriptsize{\textsf{ASC}}};
\draw [ultra thick, dashed] (7,11.15) -- (5.8,11.15);
\draw [ultra thick, dashed,-{stealth[scale=2]}] (5.8,11.15) -- (5.8,10.7);

\draw [ultra thick,-{stealth[scale=2]}] (0,9.25) -- (1.4,9.25);
\node [above] at (1.5,9.35) {\scriptsize{\textsf{Inlet}}};
\draw[fill] (1.5,9.25) circle [radius=0.1];
\draw [ultra thick] (1.5,9.25) -- (1.5,1.25);

\draw [ultra thick,-{stealth[scale=2]}] (9.1,1) -- (10,1);
\node [below] at (9,0.9) {\scriptsize{\textsf{Outlet}}};
\draw[fill] (9,1) circle [radius=0.1];
\draw [ultra thick] (9,9) -- (9,1);

\node[ultra thick,
    trapezium, 
    draw,
    rotate=-90,
    trapezium stretches = true,
    minimum width = 1.2cm, 
    minimum height = 0.7cm
    ]
    at (5.5,5) {};

\node at (5.5,5) {\scriptsize{\textsf{C2}}};

\draw (3.2,3.5) rectangle (3.9,2.8);
\node at (3.55,3.1) {\scriptsize{\textsf{FC}}};

\draw [dashed] (7,7.5) rectangle (8.1,6.8);
\node at (7.5,7.15) {\scriptsize{\textsf{ASC}}};
\draw [ultra thick, dashed] (7,7.15) -- (5.8,7.15);
\draw [ultra thick, dashed,-{stealth[scale=2]}] (5.8,7.15) -- (5.8,6.7);

\draw [rounded corners] (3.0,4.9) rectangle (4.0,4.2);
\node at (3.5,4.6) {\scriptsize{\textsf{D}}};
\node [right] at (3.55,3.8) {$\omega_{2}$};
\node [above] at (4.5,4.5) {$\tau_{2}$};

\node[ultra thick,
    trapezium, 
    draw,
    rotate=-90,
    trapezium stretches = true,
    minimum width = 1.2cm, 
    minimum height = 0.7cm
    ]
    at (5.5,1) {};

\node at (5.5,1) {\scriptsize{\textsf{C3}}};

\draw (3.2,-0.5) rectangle (3.9,-1.2);
\node at (3.55,-0.9) {\scriptsize{\textsf{FC}}};

\draw [dashed] (7,3.5) rectangle (8.1,2.8);
\node at (7.5,3.15) {\scriptsize{\textsf{ASC}}};
\draw [ultra thick, dashed] (7,3.15) -- (5.8,3.15);
\draw [ultra thick, dashed,-{stealth[scale=2]}] (5.8,3.15) -- (5.8,2.7);

\draw [rounded corners] (3.0,0.9) rectangle (4.0,0.2);
\node at (3.5,0.6) {\scriptsize{\textsf{D}}};
\node [right] at (3.55,-0.2) {$\omega_{3}$};
\node [above] at (4.5,0.5) {$\tau_{3}$};

\node [above] at (2.5,7) {${m}_{1}$};
\node [above] at (2.5,3) {${m}_{2}$};
\node [above] at (2.5,-1) {${m}_{3}$};
\end{tikzpicture}
\caption{A compressor station with three compressors C1, C2, C3, adapted from \cite{Application_Ahmed2021}. Each compressor operates at a speed $\omega_i$, $i=1,2,3$, received from a drive (D) that ensures the torque $\tau_i$ from a flow controller (FC) is achieved. The load-sharing block (LS) assigns desired flows $m_i$ to each compressor, which then responds with $m_i^c$, according to the inlet and outlet conditions. Each compressor is protected by a surge controller (ASC). Control signals are represented by dashed lines and gas flows are represented by solid lines} \label{fig:CompressorStation}
\end{figure}
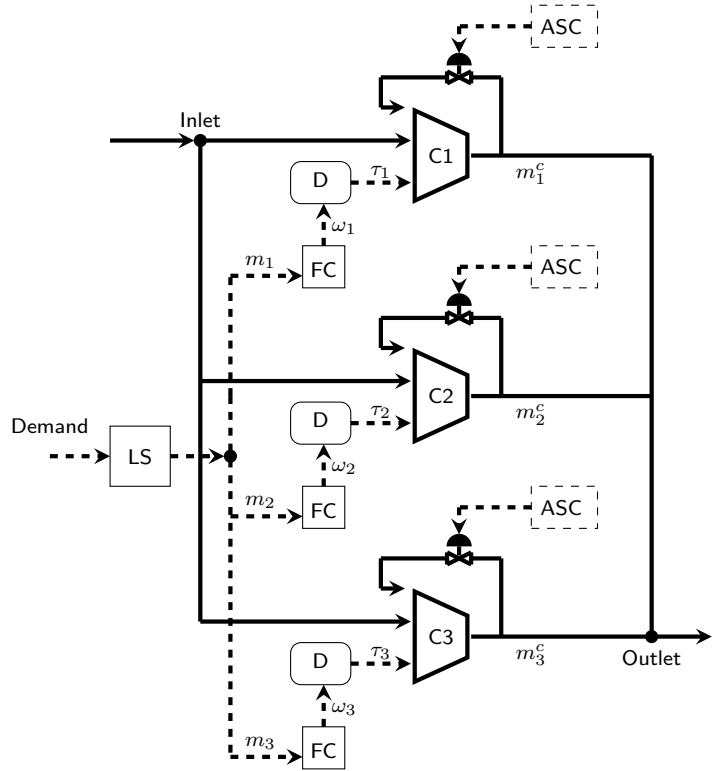

The compressor station used in this paper is shown in Fig. \ref{fig:CompressorStation} and was adapted from \cite{Application_Ahmed2021}. The station includes three centrifugal compressors, C1, C2, C3, working in parallel. Each compressor operates at a speed $\omega_i$, $i=1,2,3$, provided by a drive (D) that ensures the correct torque $\tau_i$ from a flow controller (FC) is achieved. Each compressor reacts by providing the flow $m_i^c$. Each compressor is also protected by a surge controller (ASC). Control signals are represented by dashed lines and gas flows are represented by solid lines. Each compressor has its own characteristics which are used to calculate the power consumption. The load-sharing block (LS), which is the focus of this paper, ensures that the demand is satisfied by assigning desired flows $m_i$ to each compressor, according to the inlet and outlet conditions. 

The surge controllers for each compressors ASC operate on much faster timescales than the load-sharing system. The flow controllers FC were tuned to ensure that the mass flow for a given demand is reached before the demand changes. Detailed description of both control loops was provided by \cite{Experimental_Cortinovis2015}. The current paper uses proportional-integral controllers with parameters from \cite{Real_Milosavljevic2020}.

\subsection{Compressor characteristics}
Compressor characteristics express the relationship between pressure ratio, compressor speed, mass flow rate, and compressor efficiency.  

\subsubsection{Compressor map}
The compressor map for the $i$-th compressor captures the relationship between pressure ratio $\Pi_i$, compressor speed $\omega_i$, mass flow rate $m_i^c$, and compressor efficiency $\eta_i$. Normally, the nominal map is provided by the manufacturer. 

\begin{figure}[tbp]
\psfrag{pressure}[][]{\scriptsize{\textsf{Compressor pressure ratio}}}
\psfrag{flow}[][]{\scriptsize{\textsf{Mass flow [kg s$^{-1}$]}}}
\psfrag{40}[][]{\scriptsize{\textsf{40}}}
\psfrag{50}[][]{\scriptsize{\textsf{50}}}
\psfrag{60}[][]{\scriptsize{\textsf{60}}}
\psfrag{70}[][]{\scriptsize{\textsf{70}}}
\psfrag{80}[][]{\scriptsize{\textsf{80}}}
\psfrag{90}[][]{\scriptsize{\textsf{90}}}
\psfrag{100}[][]{\scriptsize{\textsf{100}}}
\psfrag{110}[][]{\scriptsize{\textsf{110}}}
\psfrag{120}[][]{\scriptsize{\textsf{120}}}
\psfrag{130}[][]{\scriptsize{\textsf{130}}}
\psfrag{140}[][]{\scriptsize{\textsf{140}}}
\psfrag{160}[][]{\scriptsize{\textsf{160}}}
\psfrag{180}[][]{\scriptsize{\textsf{180}}}
\psfrag{200}[][]{\scriptsize{\textsf{200}}}
\psfrag{220}[][]{\scriptsize{\textsf{220}}}
\psfrag{240}[][]{\scriptsize{\textsf{240}}}

\psfrag{4}[][]{\scriptsize{\textsf{4}}}
\psfrag{3.5}[][]{\scriptsize{\textsf{3.5}}}
\psfrag{3}[][]{\scriptsize{\textsf{3}}}
\psfrag{2.5}[][]{\scriptsize{\textsf{2.5}}}
\psfrag{2}[][]{\scriptsize{\textsf{2}}}
\psfrag{1.5}[][]{\scriptsize{\textsf{1.5}}}

\psfrag{7318}[][]{\tiny{\textsf{7318}}}
\psfrag{6970}[][]{\tiny{\textsf{6970}}}
\psfrag{6621}[][]{\tiny{\textsf{6621}}}
\psfrag{6442}[][]{\tiny{\textsf{6442}}}
\psfrag{6273}[][]{\tiny{\textsf{6273}}}
\psfrag{5925}[][]{\tiny{\textsf{5925}}}
\psfrag{5576}[][]{\tiny{\textsf{5576}}}
\psfrag{5227}[][]{\tiny{\textsf{5227}}}
\psfrag{5088}[][]{\tiny{\textsf{5088}}}
\psfrag{4740}[][]{\tiny{\textsf{4740}}}

\centering
\includegraphics[width=0.8\textwidth]{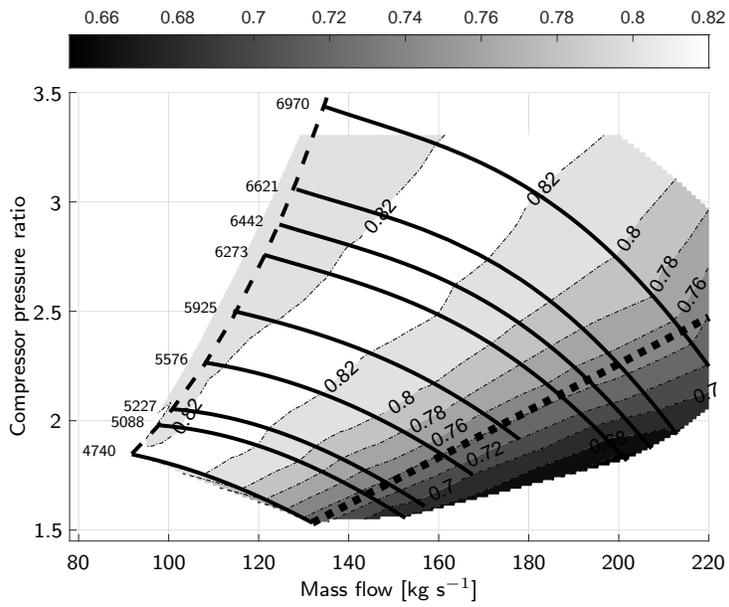}
\caption[A compressor map]{An example of compressor map with speed lines (solid lines) and efficiency islands (thin dash-dotted lines), adapted from \cite{Optimum_Noersteboe2008}. The minimal speed line corresponds to 4740 rpm, the maximal speed line 6970 rpm, the surge line (dashed line), and the choke line (dotted line) show the operating range of a compressor}
\label{fig:CompressorMapClean}
\end{figure}

Figure \ref{fig:CompressorMapClean} shows an example of a compressor map, adapted from \cite{Optimum_Noersteboe2008}. The vertical axis shows the pressure ratio of a compressor, whereas the horizontal axis presents the mass flow rate through the compressor. The compressor pressure ratio denotes the ratio between the suction pressure $p_s$ and the discharge pressure $p_d$ and is shown as a function of the mass flow rate and the speed of a compressor. Thick solid black curves in Fig. \ref{fig:CompressorMapClean} are called \textit{speed lines} and show the compressor pressure ratio as a function of the mass flow for a constant speed. The efficiency of a compressor is usually shown as a function of the compressor pressure ratio and the mass flow rate and imposed on the compressor map as constant efficiency lines which form \textit{efficiency islands} (dash-dotted lines in Fig. \ref{fig:CompressorMapClean}). The operating range for a compressor is defined by: surge line corresponding to a dynamic instability leading to oscillations and flow reversal inside the compressor (dashed line), choke line, depending on the aerodynamics of the compressor (dotted line), and minimal and maximal speed lines. The flow controllers FC and the surge controllers ASC ensure that the compressors stay within their operating ranges.

\subsubsection{Compressor efficiency}
The efficiency $\eta_i$ for each compressor can be obtained from first principles if the knowledge about gas properties is available as shown by \cite{Handbook_Mokhatab2015}, or can be estimated as shown recently by \cite{Real_Gentsch2020}. In particular, isentropic efficiency is defined as the ratio between the isentropic work and the actual work needed to compress the gas from pressure $p_s$ to $p_d$. The isentropic work depends on the mass flow through the compressor $m^c$, suction temperature $T_s$, specific heat capacity $c_p$ and the ratio of specific heats for natural gas $\gamma$ \citep{Online_Cortinovis2016}:
\begin{equation}
    W_{\text{iso}}=m^cc_pT_s\left( \left(\frac{p_s}{p_d}\right)^{\frac{\gamma-1}{\gamma}} -1 \right)
\end{equation}
The actual work depends on the mass flow and the suction and discharge temperatures $T_s$, $T_d$:
\begin{equation}
    W_{\text{act}}=m^cc_p(T_d-T_s)
\end{equation}
The efficiency can then be obtained as:
\begin{equation}
    \eta=\frac{ W_{\text{iso}}}{ W_{\text{act}}}
\end{equation}

However, gas properties as well as the temperatures may be unknown. Thus, the efficiency is usually approximated as a second order polynomial function of the mass flow through a compressor, $m_i^c$ and pressure ratio $\Pi_i$ as:
\begin{equation}
\begin{aligned}
\hat{\eta}^{\text{poly}}_i(m_i^c,\Pi_i) = &{}  \alpha_0+\alpha_1m_i^c+\alpha_2\Pi_i+\alpha_3m^c_i\Pi_i+ \alpha_4{(m^c_i)}^2+\alpha_5(\Pi_i)^2
\end{aligned}
\label{eq:EfficiencyPoly}
\end{equation}
The polynomial model from \eqref{eq:EfficiencyPoly} is considered a good representation of the functional form of compressor efficiency maps \citep{Modeling_Egeland2002a}. The coefficients $\alpha_i$, $i=0,\ldots,5$ can be estimated from measured data for $m_i^c$, $\Pi_i$, $\eta_i$ \citep{Online_Cortinovis2016}. 

\subsubsection{Compressor head}
The compressor head captures the thermodynamics of a compression process \citep[Ch. 8.15]{Instrument_Liptak2005}. At a given pressure ratio $\Pi_i$, the compressor head $H$ is calculated as:
\begin{equation}
H=\frac{ZRT_1}{M_{\text{W}}\phi}\left( \Pi_i^{\phi} -1 \right)
\label{eq:CompressorHead}
\end{equation}
where $\phi=\frac{n-1}{n}$ with $n$ as the polytropic coefficient and $\Pi_i$ is the pressure ratio across the compressor. $M_{\text{W}}$ denotes the molecular weight of the gas, $R$ is the gas constant, $Z$ denotes the compressibility of the gas, and $T_1$ is the inlet temperature of the compressor. The values used in the paper are for natural gas from \cite{Real_Milosavljevic2020}.

\subsubsection{Compressor power}
The power necessary to run a compressor with efficiency $\eta_i$ at the head $H_i$ and the flow $m_i^c$ is given as:
\begin{equation}
W_i = \frac{H_im^c_i}{\eta_i}.
\label{eq:CompressorPower}
\end{equation}
The head of the compressor in \eqref{eq:CompressorPower} depends on the properties of the gas and the pressure ratio as indicated in \eqref{eq:CompressorHead}. As such, the power will depend on the operating point of a compressor. The available operating points for a single compressor are defined by the intersection of the \emph{system resistance curve} and the characteristics of a compressor \citep[Ch. 7.10]{Instrument_Liptak2005}. The resistance curve is independent from the current operating state of the compressors and is determined based on system characteristics such as the piping geometry and the properties of the processed fluid. In the current paper, the properties of the gas and the internal piping are assumed constant and the resistance curve is assumed to be linear:
\begin{equation}
    \Pi_i(m_i)=\rho_1m_i+\rho_2
    \label{eq:Resistance}
\end{equation}
where $\rho_1=0.017$ and $\rho_2=0.78$ are constant for all $i=1,2,3$. The values of $\rho_1$ and $\rho_2$ have been adapted from \cite{Experimental_Cortinovis2015}. The importance of the resistance curve is emphasised in particular for centrifugal compressors with variable speed drives which are considered in this work. Adjusting the speed according to the required demand allows mitigating the power consumption by moving the operating point along the resistance curve \citep[Ch. 7.10]{Instrument_Liptak2005}.

\subsection{Optimal load sharing}
\label{sec:OptimalLS}
 In this paper, the optimal load sharing problem is formulated as minimisation of power consumption $W$ of the whole station, following \cite{Influence_Zagorowska2019}:
\begin{equation}
W(m_1^c,m_2^c,m_3^c) = W_1(m_1^c)+W_2(m_2^c)+W_3(m_3^c)
\label{eq:MinPower}
\end{equation}
where $W_i(m^c_i)$ is calculated using \eqref{eq:CompressorPower}.
The demand constraint is added to satisfy the external demand $M$:
\begin{equation}
m^c_1+m^c_2+m^c_3 = M
\label{eq:DemandConstraint}
\end{equation}
In the current paper we assume that the demand $M$ is such that the compressor station can always satisfy it. Alternative approaches were reviewed for instance by \cite{Optimal_Xenos2015}.

\section{Online adaptation of compressor efficiency map using Gaussian process regression}
\label{sec:GPefficiency}
\cite{Real_Milosavljevic2020} indicated that efficiency maps are the main element of compressor models characterised by uncertainty. Thus, the efficiency maps in this work are adapted using Gaussian process regression as proposed by \cite{Application_Ahmed2021}. Gaussian process regression has been shown to work well with a limited number of measurements \citep{Gaussian_Rasmussen2006} and its performance has already been validated for centrifugal compressors \citep{Data_Korkmaz2022}. 

\subsection{Efficiency error approximation}
In this paper, the difference between the real efficiency and the approximation is given as:
\begin{equation}
\Delta^k_i(m_i^{ck},\Pi_i^k) = \eta^k_i-\hat{\eta}^{\text{poly},k}_i(m_i^{ck},\Pi_i^k)
\label{eq:Error}
\end{equation}
where $\eta^k_i$ is the real efficiency measured from the $i$-th compressor at a time instant $k$, $\hat{\eta}^{\text{poly},k}_i$ is the approximated efficiency evaluated at the measured pressure ratio $\Pi_i^k$ and mass flow $m_i^{ck}$ using \eqref{eq:EfficiencyPoly}. Using \eqref{eq:Error} allows preservation of functional form of efficiency maps from \eqref{eq:EfficiencyPoly}. Preserving the functional form provides additional information for the approximation of efficiency, resulting in learning the correct efficiency more quickly \citep{Application_Ahmed2021}.

\subsection{Gaussian process regression}
\label{sec:GPs}
Following \cite[Ch. 2]{Gaussian_Rasmussen2006}, the error $\Delta_i$ from \eqref{eq:Error} in the current paper is modelled as a function of the mass flow $m_i^c$ and pressure ratio $\Pi_i$:
\begin{equation}
\Delta_i(m^c_i,\Pi_i) = g(m^c_i,\Pi_i)+\varepsilon
\label{eq:InitialDeltaModel}
\end{equation}
where $\varepsilon$ is Gaussian noise with zero mean and variance $\sigma^2$. The indicator $k$ has been omitted for simplicity. The function $g(m^c_i,\Pi_i)$ describes how the model depends on the mass flow and the pressure ratio:
\begin{equation}
g(m^c_i,\Pi_i) = \beta + f(m^c_i,\Pi_i)
\end{equation}
where $\beta$ is a coefficient and $f(\cdot,\cdot)$ is a Gaussian process with zero mean and covariance function $\kappa_{\theta}((m^c_i,\Pi_i),({m^c_i}',\Pi_i'))$ parametrised by a vector of hyperparameters $\theta$:
\begin{equation}
f(m^c_i,\Pi_i)\sim GP(0,\kappa_{\theta}((m^c_i,\Pi_i),({m^c_i}',\Pi_i'))).
\end{equation}
From \cite{Gaussian_Rasmussen2006}, we obtain that the function $g(\cdot)$ is also a Gaussian process:
\begin{equation}
    g(m^c_i,\Pi_i)\sim GP(\beta,\kappa_{\theta}((m^c_i,\Pi_i),({m^c_i}',\Pi_i'))).
    \label{eq:GPforg}
\end{equation}
Using \eqref{eq:GPforg} and \eqref{eq:InitialDeltaModel}, we can model the error $\Delta_i$ as:
\begin{equation}
    \Delta_i(m^c_i,\Pi_i)\sim GP(\beta,\kappa_{\theta}((m^c_i,\Pi_i),({m^c_i}',\Pi_i'))+\sigma_n^2\cdot\delta_{ii'}).
    \label{eq:DeltaGaussian}
\end{equation}
where $\delta_{ii'}=1$ if $i=i'$ and zero otherwise \citep{Gaussian_Rasmussen2003}.

\subsubsection{Covariance function}
\label{sec:GPRcovariance}

To use \eqref{eq:DeltaGaussian} for prediction, it is necessary to specify the covariance function $\kappa_{\theta}(\cdot)$ parametrised by $\theta$. In this paper, we chose a squared exponential covariance function:
\begin{equation}
    \kappa_{\theta}(x,x')=\theta_f^2\exp\left( -\frac{(x-x')^{\T}(x-x')}{2\theta_l} \right)
    \label{eq:SquaredExp}
\end{equation}
where $\theta=[\theta_f^2,\theta_l]^{\T}$ with $\theta_f^2$ as the signal variance and $\theta_l$ as the characteristic length scale. The values of $\theta_f^2$ and $\theta_l$ are estimated from data, as will be described in Section \ref{sec:GPRestimation}. For simplification, in \eqref{eq:SquaredExp} we have $x=(m^c_i,\Pi_i)$ and $x'=({m^c_i}',\Pi_i')$.

The covariance function \eqref{eq:SquaredExp} indicates that points $x$, $x'$ close to each other will have covariance close to $\theta_f^2$ whereas points far from each other will have covariance going to zero \citep{Gaussian_Rasmussen2006}. This property of the covariance \eqref{eq:SquaredExp} allows us to model the error $\Delta_i$ as slowly changing with $m^c_i,\Pi_i$ because a compressor should have similar efficiency for small changes of the mass flow and the pressure \cite[Ch. 5]{Optimum_Noersteboe2008}. The covariance function \eqref{eq:SquaredExp} also ensures differentiability of the estimated error, which will be required for OFO in Section \ref{sec:FO}.

\subsubsection{Estimation of parameters}
\label{sec:GPRestimation}
Gaussian process regression estimates the parameters, $\hat{\beta}$, $\hat{\theta}$, and $\hat{\sigma}^2$ corresponding to unknown ${\beta}$, ${\theta}$, and ${\sigma}^2$ using $k$ measured values ${m^c_i}^j$, $\Pi_i^j$, and $\Delta_i^j$, $j=1,\ldots,k$. The estimated parameters are obtained from maximisation of a marginal likelihood function:
\begin{equation}
\begin{aligned}
    \hat{\beta},\hat{\theta},\hat{\sigma}_n^2=&{}\argmax_{\beta,\theta,\sigma_n^2} -\frac{1}{2}( \boldsymbol{\Delta}_i-\beta I_k)^{\T}(K_{\theta}(\mathbf{X}_i,\mathbf{X}_i')+\sigma_n^2I_{k^2})^{-1}( \boldsymbol{\Delta}_i-\beta I_k)\\
    -&{}\frac{k}{2}\log 2\pi-\frac{1}{2}\log |K_{\theta}(\mathbf{X}_i,\mathbf{X}_i')+\sigma_n^2I_{k^2}|
    \end{aligned}
    \label{eq:MaxLikelihood}
\end{equation}
where $\boldsymbol{\Delta}_i=[\Delta_i^j]_{j=1,\ldots,k}$, $\mathbf{X}_i=[x^j_i]_{j=1,\ldots,k}$ with $x_i^j = ({m^c_i}^j, {\Pi_i^j})$, and 
\begin{equation}
K_{\theta}(\mathbf{X}_i,\mathbf{X}_i') = \begin{bmatrix}
\kappa_{\theta}(x_i^q,x_i^r)
\end{bmatrix}_{q,r=1\ldots,k}
\end{equation}
Finally, $I_{k^2}$ denotes an identity matrix of size $k\times k$ and $I_k$ denotes a vector of size $k$ \citep{Gaussian_Rasmussen2003}.

\subsubsection{Prediction}
\label{sec:GPRprediction}
To predict the error corresponding to a new point $x_i^{\text{new}}=[m_i^{\text{c,new}},\Pi_i^{\text{new}}]$ we use the fact that the joined distribution of known values $\boldsymbol{\Delta}_i$ and the unknown error $\Delta_i^{\text{new}}$ is a normal distribution \citep{Gaussian_Rasmussen2003}:
\begin{equation}
    \begin{bmatrix}
    \boldsymbol{\Delta}_i\\
    \Delta_i^{\text{new}}
    \end{bmatrix}\sim
    \mathcal{N}\left(\begin{bmatrix}
    \beta I_k\\
    \beta
    \end{bmatrix},\begin{bmatrix}
    K_{\theta}(\mathbf{X}_i,\mathbf{X}_i')+\sigma^2_nI_{k}& K_{\theta}(x_i^{\text{new}},\mathbf{X}_i')\\
    K_{\theta}(x_i^{\text{new}},\mathbf{X}_i')^{\T} &  K_{\theta}(x_i^{\text{new}},x_i^{\text{new}})
    \end{bmatrix}\right)
\end{equation}
The conditional distribution of $\Delta_i^{\text{new}}$ given the known values $\boldsymbol{\Delta}_i$ is also a normal distribution:
\begin{equation}
    {\Delta}_i^{\text{new}}|\boldsymbol{\Delta}_i\sim\mathcal{N}\left(\mu^{\text{new}},k^{\text{new}} \right)
\end{equation}
where
    \begin{align}
    \mu^{\text{new}}=&{}\beta+ K_{\theta}(x_i^{\text{new}},\mathbf{X}_i')^{\T}(K_{\theta}(\mathbf{X}_i,\mathbf{X}_i')+\sigma^2_nI_{k})^{-1}(\boldsymbol{\Delta}_i-I_{k}\beta)\label{eq:Newmean}\\
    k^{\text{new}}=&{} K_{\theta}(x_i^{\text{new}},x_i^{\text{new}})-K_{\theta}(x_i^{\text{new}},\mathbf{X}_i')^{\T}(K_{\theta}(\mathbf{X}_i,\mathbf{X}_i')+\sigma^2_nI_{k})^{-1}K_{\theta}(x_i^{\text{new}},\mathbf{X}_i')
    \end{align}

\subsection{Estimation of compressor efficiency using Gaussian process regression}

\subsubsection{Error estimation}
\label{sec:ErrorEstimation}
Estimating the error in compressor efficiency requires the information about the new mass flow $m_i^{\text{new}}$ and the pressure ratio $\Pi_i^{\text{new}}$. We note that the new pressure ratio is a function of the new mass flow, connected by the resistance curve from \eqref{eq:Resistance}. Using the estimated parameters $\hat{\beta}$, $\hat{\theta}$, and $\hat{\sigma}^2$, the error $\hat{\Delta}_i^{\text{new}}$ corresponding to a new point $x_i^{\text{new}}= [m_i^{\text{new}}, \Pi_i^{\text{new}}]$ is then calculated from \eqref{eq:Newmean} as:
\begin{equation}
\hat{\Delta}_i^{\text{new}}=\hat{\beta}+K_{\hat{\theta}}(x_i^{\text{new}},\mathbf{X}_i)^{\T}(K_{\hat{\theta}}(\mathbf{X}_i,\mathbf{X}_i'))+\hat{\sigma}^2_nI_{k})^{-1}(\boldsymbol{\Delta}_i-I_{k}\hat{\beta})
\label{eq:EstimatedError}
\end{equation}
where $K_{\hat{\theta}}(x_i^{\text{new}},\mathbf{X}_i)$ evaluates the covariance between the new point and existing data \citep{Gaussian_Rasmussen2003}. 
 
The estimate $\hat{\Delta}_i^{\text{new}}$ together with the polynomial approximations from \eqref{eq:EfficiencyPoly} is then used to approximate the efficiency $\hat{\eta}_i^{\text{new}}$ as
\begin{equation}
\hat{\eta}_i^{\text{new}}(m_i^{c,new},\Pi_i^{new}) = \hat{\eta}^{\text{poly},\text{new}}_i(m_i^{c,new},\Pi_i^{new})+ \hat{\Delta}_i^{\text{new}}(m_i^{c,new},\Pi_i^{new}).
\label{eq:EstimatedEfficiency}
\end{equation}
The estimated efficiency $\hat{\eta}_i^{\text{new}}$ is then used to calculate power consumption of the $i$-th compressors using \eqref{eq:CompressorPower} in Section \ref{sec:FOcompressors}.

\subsubsection{Online adaptation}
The prediction of the error from Section \ref{sec:ErrorEstimation} at a new point $x_i^{\text{new}}=[m_i^{\text{new}}, \Pi_i^{\text{new}}]$ requires $k$ known values of efficiency $\eta^k_i$ from the $i$-th compressor corresponding to the past measured pressure ratio $\Pi_i^k$ and mass flow $m_i^k$. However, if the new point $x_i^{\text{new}}$ is far in the sense of the Euclidean norm from the measured points $x_i^1,\ldots,x_i^k$, then \eqref{eq:EstimatedError} will become constant:
\begin{equation}
    \hat{\Delta}_i^{\text{new}}\approx\hat{\beta}
    \label{eq:ConstantError}
\end{equation}
because $K_{\hat{\theta}}(x_i^{\text{new}},\mathbf{X}_i)\approx 0$. The constant error from \eqref{eq:ConstantError} would correspond to a mismatch constant for all mass flows and pressures in \eqref{eq:Error}. \cite{Online_Cortinovis2016} indicated that a constant mismatch is unlikely and adaptation is required to enable better prediction of the overall efficiency.

In this paper, we propose to adapt the efficiency estimate by applying Gaussian process regression from Section \ref{sec:GPs} in an online way. By iteratively extending the data used for estimation of parameters of the Gaussian process, we can improve the prediction for the error. At the same time, we reuse existing ways of estimating the error if no new information is available. Reusing existing information allows us to reduce the computational cost of fitting Gaussian processes to large datasets, which can be significant \citep{Gaussian_Rasmussen2006}.

The procedure was adapted from \cite{Application_Ahmed2021} and is shown in Algorithm \ref{alg:AdaptationWithGPR}. We assume that the polynomial model \eqref{eq:EfficiencyPoly} is given, the sets of past measurements $\mathbf{X}$, $\boldsymbol{\Delta}$ are available, the corresponding function $\hat{\Delta}(\cdot,\cdot)$ for estimating the error is known and has covariance function $\kappa_{\theta}(\cdot,\cdot)$. When a new measurement $m^{c,\text{new}}_i$, $\Pi_i^{\text{new}}$, $\eta^{\text{new}}_i$ comes in, the algorithm computes $\Delta^k$ from \eqref{eq:Error}. If the existing sets $\mathbf{X}$, $\boldsymbol{\Delta}$ are empty, or if they only contain the same points as the new measurement, the algorithm uses the existing function $\hat{\Delta}(\cdot,\cdot)$ for estimating the error (lines two to four). In a similar way, if the existing sets already contain the new measurement, the algorithm also uses the existing function $\hat{\Delta}(\cdot,\cdot)$ (lines six and seven). The estimation of parameters for a new Gaussian process is only triggered if the existing sets do not contain the new measurement (lines nine to 12). The algorithm then returns the function $\hat{\Delta}(\cdot,\cdot)$ that allows prediction of the error, as well as the updated sets with measurements $\mathbf{X}_{\text{new}}$ and $\boldsymbol{\Delta}_{\text{new}}$. The function $\hat{\Delta}(\cdot,\cdot)$ can then be used in Online Feedback Optimization, as described in Section \ref{sec:FOcompressors}.

\begin{algorithm}[t]
\SetAlgoLined
  \KwInput{Initial model \eqref{eq:EfficiencyPoly}, new measurement $m_i^{c,\text{new}}$, $\Pi_i^{c,\text{new}}$, $\eta_i^{\text{new}}$, sets of past measurements $\mathbf{X}$, $\boldsymbol{\Delta}$, and the corresponding function $\hat{\Delta}(\cdot,\cdot)$, covariance function $\kappa_{\theta}(\cdot,\cdot)$}
  \KwOutput{Function to estimate $\hat{\Delta}(m^c_i,\Pi_i)$, updated sets $\mathbf{X}_{\text{new}}$, $\boldsymbol{\Delta}_{\text{new}}$}
   Evaluate $\Delta^k_i$ from \eqref{eq:Error}
   
\eIf{$|\mathbf{X}\setminus \lbrace(m_i^{c,\text{new}},\Pi_i^{c,\text{new}})\rbrace|=0$  {\bf and} $|\boldsymbol{\Delta}\setminus \lbrace \Delta^k_i \rbrace|=0$}{
   Set the function $\hat{\Delta}_0: (m,\Pi)\rightarrow \hat{\Delta}(m,\Pi)$
   
   Set $\mathbf{X}_k\leftarrow \lbrace(m_i^{c,\text{new}},\Pi_i^{\text{new}})\rbrace$ and $\boldsymbol{\Delta}_k\leftarrow \Delta^k_i$

   }{
   \eIf{$\mathbf{X}\cap \lbrace(m_i^{c,\text{new}},\Pi_i^{\text{new}})\rbrace\neq\varnothing$ {\bf or} $\boldsymbol{\Delta}\cap \lbrace\Delta^k\rbrace\neq\varnothing$ }{
   Set the function $\hat{\Delta}_0: (m,\Pi)\rightarrow \hat{\Delta}(m,\Pi)$

   }{
   

   Set $k\leftarrow |\mathbf{X}|+1$
   
   Set $\mathbf{X}_k\leftarrow \mathbf{X}\cup \lbrace(m_i^{c,\text{new}},\Pi_i^{\text{new}})\rbrace$ and  $\boldsymbol{\Delta}_k\leftarrow \boldsymbol{\Delta} \cup \lbrace\Delta^k\rbrace$
   
   For the chosen covariance function $\kappa_{\theta}(\cdot,\cdot)$ and the sets $\mathbf{X}_k$,    $\boldsymbol{\Delta}_k$ solve \eqref{eq:MaxLikelihood} to obtain $\hat{\beta}, \hat{\theta},\hat{\sigma}_n^2$
   
   Define function $\hat{\Delta}_0: (m,\Pi)\rightarrow \hat{\beta}+ K_{\hat{\theta}}((m,\Pi),\mathbf{X}_k)^{\T}(K_{\hat{\theta}}(\mathbf{X}_k,\mathbf{X}_k')+ \hat{\sigma}^2_n I_{k})^{-1} (\boldsymbol{\Delta}_k-I_{k}\hat{\beta}$)

   }

   }
   Set the function $\hat{\Delta}\leftarrow \hat{\Delta}_0$
   
   Set $\mathbf{X}_{\text{new}}\leftarrow \mathbf{X}_k$ and $\boldsymbol{\Delta}_{\text{new}}\leftarrow \boldsymbol{\Delta}_k$

 \caption{Adaptive estimation of compressor efficiency using Gaussian process regression adapted from \cite{Application_Ahmed2021}}
 \label{alg:AdaptationWithGPR}
\end{algorithm}

\section{Online Feedback Optimization with model adaptation}
\label{sec:FOcompressors}

Online Feedback Optimization has been successfully validated in optimization of power grids by \cite{Experimental_Ortmann2020} where the authors have shown that OFO has good tracking performance for time-varying reference signal. The ability to follow time-varying reference signal makes OFO useful for compressor stations where satisfaction of demand is of importance.

\subsection{Online Feedback Optimization}
\label{sec:FO}
The main idea of \ac{OFO} is to treat optimization algorithms as dynamic systems. The dynamic system representing the optimization algorithm is then connected in a closed loop with the controlled system with inputs $u$ and outputs $y$. \ac{OFO} iteratively updates the input of a system $u$ to make the system converge to a local optimum of an optimization problem. In particular, \ac{OFO} makes use of measured outputs instead of full models to solve the optimization problem \citep{Optimization_Hauswirth2021}. 

The optimization problem in \ac{OFO} is formulated as:
\begin{subequations} \label{eqn:ProblemStatement}
\begin{align}
\min_{u,y}& \quad \Phi(u,y)
    \label{eqn:CostFcn}\\
\text{subject to }    & y=h(u)\label{eq:mapping}\\
    &u\in\mathcal{U}, y\in\mathcal{Y}
\end{align}
\end{subequations}
where $\Phi:\R^p\times\R^n\rightarrow\R$ is a continuously differentiable cost function, $h:\R^p\rightarrow\R^n$ is a continuously differentiable nonlinear output mapping in steady state, and $\mathcal{U}$ and $\mathcal{Y}$ describe the constraints on the inputs and outputs, respectively:
\begin{center}
$\mathcal{U}=\lbrace u\in\R^p:Au\leq b \rbrace$ and  $\mathcal{Y}=\lbrace y\in\R^n:Cy\leq d \rbrace$  
\end{center}
where $A\in \R^{q\times p}$, $b\in\R^q$, $C\in\R^{l\times n}$, and $d\in\R^l$ are constant matrices \citep{Non_Haeberle2020}.

The \ac{OFO} controller used in this paper was proposed by \cite{Non_Haeberle2020}. It is a discrete integral feedback controller with constant step size \(\nu>0\) 
\begin{align}\label{eqn:Verena_feedback}
    u^{k+1} = u^k + \nu\widehat{\sigma}_\nu (u^k,y^k) \quad \text{with } y^k = h(u^k),
\end{align} 
where $y^k = h(u^k)$ is the measured system output at time $k$, and \(\widehat{\sigma}_\nu (u^k,y^k)\) is the minimizer of the constrained optimization problem
\begin{subequations} \label{eqn:Verena_opt}
\begin{align}
    \widehat{\sigma}_\nu (u,y) = &\argmin_{w\in\mathbb{R}^p}\left|\left| w + H^\top(u)\nabla\Phi^\top(u,y)\right|\right|^2
    \label{eqn:Verena_sigma}\\
    &\text{subject to}\quad A\left(u^k+\nu w\right)\leq b\\
    &\qquad\qquad\quad C \left(y^k+\nu\nabla h(u^k) w\right)\leq d.
\end{align}
\end{subequations}
where $w$ is an auxiliary decision variable of the same size as the system input $u$. For space reasons, the superscript $(\cdot)^k$ is dropped in~(\ref{eqn:Verena_sigma}). The space of all feasible inputs is \(\mathcal{U} = \{u \in \mathbb{R}^p | Au\leq b\}\), and \(H(u^k)^\top = \left[\mathbb{I}_p \ \nabla h(u^k)^\top\right]\), with $\nabla h(u^k)^\top$ is called \emph{input-output sensitivity}. The matrix $\mathbb{I}_p$ is an identity matrix of size $p\times p$. The gradient of the objective function $\nabla\Phi(u,y)$ is defined as
\begin{equation}
    \nabla\Phi(u,y)=\begin{bmatrix}
    \frac{\partial \Phi}{\partial u} & \frac{\partial \Phi}{\partial y}
    \end{bmatrix}=\begin{bmatrix}
    \frac{\partial \Phi}{\partial u_1} & \ldots & \frac{\partial \Phi}{\partial u_p} & \frac{\partial \Phi}{\partial y_1} & \ldots & \frac{\partial \Phi}{\partial y_n}
    \end{bmatrix}
    \label{eq:GradientDef}
\end{equation}
Only the gradient of the objective function $\nabla \Phi(u,y)$ in \eqref{eqn:Verena_sigma} and the input-output sensitivity $\nabla h(u^k)^\top$ are necessary for OFO. In particular, $\nabla h(u^k)^\top$ can be computed analytically if the mapping $h(u)$ is known, or estimated numerically. The computation of gradients in this work will be described in Section \ref{sec:PowerEstim}.  

The optimization problem \eqref{eqn:Verena_opt} is quadratic and convex, and therefore easy to solve.

\subsection{Online Feedback Optimization for compressor station}
\subsubsection{Input-output formulation} \ac{OFO} from Section \ref{sec:FO} requires a formulation in terms of inputs and outputs. The inputs of the load sharing problem from Section \ref{sec:OptimalLS} are the target mass flows $m_i$ assigned to each compressor. The outputs are the measured mass flows $m^c_{i}$ provided by each compressor and the individual power consumption $W_i$. Introducing $u=[m_1,m_2,m_3]^{\top}$ and $y=[m^c_{1}, m^c_{2}, m^c_{3}, W_1,W_2,W_3]^{\top}$, it is obtained that the function $h(u)$ from \eqref{eq:mapping} describing the mapping from inputs $u$ to outputs $y$ is:
\begin{equation}
h(u)=[
m^c_1(u)\;m^c_2(u)\;m^c_3(u)\;W_1(u)\;W_2(u)\;W_3(u)
]^{\top}
\label{eq:compressormapping}
\end{equation} 
where $m^c_i$ denote the outlet flow of each compressor if the load-sharing block assigns the flow $m_i$. In steady state the two flows are equal, $m^c_i=m_i$. The objective function from \eqref{eq:MinPower} can be then written as:
\begin{equation}
\Phi(u,y) = \begin{bmatrix}
0&0&0&1&1&1
\end{bmatrix}y.
\label{eq:RewrittenObj}
\end{equation}
The constraints on the inputs $m_i$ can be written as:
\begin{equation}
m_i^{\min}\leq m_i\leq m_i^{\max}
\label{eq:RewrittenCstr}
\end{equation}
where $m_i^{\min}$ and $m_i^{\max}$ are the minimal and maximal flow allowed for the $i$-th compressor. The constraint \eqref{eq:RewrittenCstr} corresponds to choosing the operating range for each compressor as a subset of compressor characteristic bounded by $m_i^{\min}$, $m_i^{\max}$, $N_i^{\min}$, and $N_i^{\max}$. Thanks to the low-level controllers (FC and ASC in Fig. \ref{fig:CompressorStation}) we can assume that this operating range is feasible. 

The formulation of \ac{OFO} presented in Section \ref{sec:FO} assumes that the constraints on the outputs are also formulated as inequality constraints. To apply \ac{OFO} to a compressor station with a demand constraint from \eqref{eq:DemandConstraint}, the equality constraint \eqref{eq:DemandConstraint} was written as two inequality constraints:
\begin{equation}
M\leq m_1+m_2+m_3\leq M
\label{eq:DemandRelaxation}
\end{equation}
The formulation from \eqref{eq:DemandRelaxation} indicates that the demand must always be satisfied, otherwise the problem will be infeasible. To ensure the feasibility, we assume that the demand is within the limits of the system. This assumption is justified if we consider that the demand required from the station is provided by external operators who are aware of the limits of the station.

\subsubsection{Quadratic optimization problem} 
The quadratic optimization problem from \eqref{eqn:Verena_opt} for the load-sharing problem with objective function \eqref{eq:RewrittenObj} becomes:
\begin{subequations} \label{eqn:compressor_opt}
\begin{align}
    \widehat{\sigma}_\nu (u,y) =&{} \argmin_{w\in\mathbb{R}^3}\frac{1}{2}w^{\top}w+\nabla \Phi^{\top}(u,y)H^{\top}w	
    \label{eqn:compressorsigma}\\
    &\text{subject to}\quad A\left(u^k+\nu w\right)\leq b\label{eq:compconstraints}\\
    &\qquad\qquad\quad C \left(y^k+\nu\nabla h(u^k) w\right)\leq d\label{eqn:Cdmatrices}
\end{align}
\end{subequations}
where $h(u)$ is given by \eqref{eq:compressormapping} and:
\begin{equation}
    H=\begin{bmatrix}
1&0&0&1&0&0&\frac{\d W_1}{\d m_1}&0&0\\
0&1&0&0&1&0&0&\frac{\d W_2}{\d m_2}&0\\
0&0&1&0&0&1&0&0&\frac{\d W_2}{\d m_2}\\
\end{bmatrix}
\label{eq:Hmatrix}
\end{equation}
and from \eqref{eq:GradientDef} where $p=3$ and $n=6$:
\begin{equation*}
    \nabla \Phi(u,y)=\begin{bmatrix}
0&0&0&0&0&0&1&1&1
\end{bmatrix}
\end{equation*}

The constraints \eqref{eq:compconstraints} are defined by matrices $A$ and $b$:
\begin{center}
$A=\begin{bmatrix}
1& 0& 0\\
        -1& 0& 0\\
        0& 1 &0\\
        0 &-1& 0\\
        0& 0& 1\\
        0& 0& -1
\end{bmatrix}$, $b=\begin{bmatrix} m_1^{\max}\\ -m_1^{\min}\\ m_2^{\max}\\ -m_2^{\min}\\ m_3^{\max}\\ -m_3^{\min}\end{bmatrix}$. 
\end{center}
The optimization problem \eqref{eqn:compressor_opt} ensures that the constraints on the inputs and outputs are satisfied in steady state. Possible transient behaviour of a compressor is handled by the flow controllers FC and the surge controllers ASC (Fig. \ref{fig:CompressorStation}) that ensure that the compressors stay within their operating ranges.

Matrices $C$ and $d$ from \eqref{eqn:Cdmatrices} correspond to the demand constraint \eqref{eq:DemandConstraint}. The inequality constraint \eqref{eq:DemandRelaxation} yields matrices:
\begin{center}
$C=\begin{bmatrix}
-1& -1& -1& 0& 0& 0\\
        1& 1 &1& 0& 0& 0
\end{bmatrix}$, $d=\begin{bmatrix}-1\\ 1\end{bmatrix}M$.
\end{center}


\subsection{Online Feedback Optimization with model adaptation}
The Online Feedback Optimization from \ref{sec:FOcompressors} is now combined with the adaptation of efficiency from Section \ref{sec:GPefficiency}.  

\subsubsection{Power estimation}
\label{sec:PowerEstim}
The power consumption for the $i$-th compressor is estimated using \eqref{eq:EstimatedEfficiency} as:
\begin{equation}
\label{eq:PoweREstimate}
\hat{W}_i=\frac{H(m_i)m_i}{\hat{\eta}_i(m_i)}.
\end{equation}
To compute the power from \eqref{eq:PoweREstimate} we evaluate the compressor head $H(m_i)$ from \eqref{eq:CompressorHead} using the resistance curves from \eqref{eq:Resistance}. Under changing system resistance conditions, the resistance curves can be updated in a data-driven way similar to the compressor efficiency maps.

The optimization problem in \ac{OFO} for a compressor station requires the evaluation of derivatives of the power estimate with respect to the mass flow, $\frac{\d\hat{W}_i}{\d m_i}$. The derivatives are approximated using forward finite differences:
\begin{equation}
\frac{\d\hat{W}_i}{\d m_i}\approx\frac{\hat{W}_i(m_i+\delta)-\hat{W}_i(m_i)}{\delta}
\end{equation}
where $\delta=1$e$-8$. To ensure differentiability of the estimate $\hat{W}_i(m_i)$, the covariance function from Section \ref{sec:GPefficiency} is chosen as a squared exponential covariance function with constant basis functions \cite[Ch. 4]{Gaussian_Rasmussen2006}. 

\subsubsection{Block diagram}
\begin{figure}[tbp]
\psfrag{Controller}[][]{\scriptsize{\textsf{OFO}}}
\psfrag{Actuator}[][]{\scriptsize{\textsf{Physics}}}
\psfrag{System}[][]{\scriptsize{\textsf{Adaptation}}}
\psfrag{Degradation}[][]{\scriptsize{\textsf{Power}}}
\psfrag{calculation}[][]{\scriptsize{\textsf{estimation}}}
\psfrag{gradient}[][]{\scriptsize{$\frac{\d\hat{W}_i}{\d m_i}\Big|_{m_i=m_i^c}$}}
\psfrag{massin}[][]{\scriptsize{$m_i$}}
\psfrag{pi}[][]{\scriptsize{$\Pi_i$,}}
\psfrag{eta}[][]{\scriptsize{$\eta_i$}}
\psfrag{External}[][]{\scriptsize{\textsf{resistance curves,}}}
\psfrag{factors}[][]{\scriptsize{\textsf{efficiency model} \eqref{eq:EfficiencyPoly}}}
\psfrag{mass}[][]{\scriptsize{$m_i^c$}}
\psfrag{demand}[][]{\scriptsize{$M$}}
\psfrag{eficiency}[][]{\scriptsize{\hspace{0.5cm}$\hat{\Delta}(m_i^c,\Pi_i)$}}

\centering
\includegraphics[width=1\textwidth]{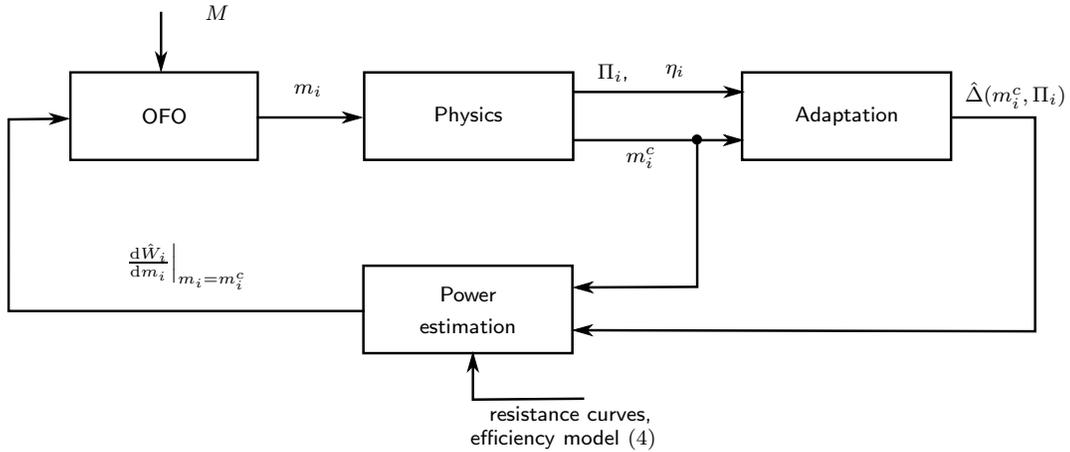}
\caption[A compressor map]{A block diagram of \ac{OFO} for the compressor station from Fig. \ref{fig:CompressorStation}}
\label{fig:BlockDiagram}
\end{figure}

Figure \ref{fig:BlockDiagram} presents a block diagram of \ac{OFO} with model adaptation. The physics of the compressor station from Fig. \ref{fig:CompressorStation} is affected by the set points $m_i$ obtained from \ac{OFO}. The pressure ratios $\Pi_i$, the true efficiencies $\eta_i$ and the corresponding mass flows $m_i^c$ from the $i$-th compressor, $i=1,2,3$ are then passed to the adaptation block described in Section \ref{sec:GPefficiency}. The estimated errors $\hat{\Delta}(m_i^c,\Pi_i)$ are used to compute the efficiencies $\hat{\eta}_i$, which are needed to estimate the power consumption of the $i$-th compressor and evaluate the derivatives in \eqref{eq:Hmatrix} described in Section \ref{sec:PowerEstim}. In the current case study, we assume the system resistance curve is available as a model and is provided to the power estimation block for calculating the pressure change for a given total station flow. The derivatives, evaluated at the measured values of the mass flows through each compressor, are then fed into the \ac{OFO} block. The \ac{OFO} block calculates the new set points $m_i$ based only on the derivatives and the current demand $M$.

\section{Results}
\label{sec:Results}
The proposed approach was analysed in four scenarios:
\begin{itemize}
\item Solving the nonlinear optimization problem assuming no mismatch (NLP) - benchmark solution
\item Using \ac{OFO} without mismatch
\item Using \ac{OFO} with mismatch and no adaptation
\item Using \ac{OFO} with mismatch and online adaptation
\end{itemize}
The optimization problem from Section \ref{sec:FOcompressors} was solved using \texttt{quadprog} in Matlab Version 9.10.0.1739362 (R2021a). Estimation of the parameters for Gaussian process regression from Section \ref{sec:GPRestimation} was performed by the function \texttt{fitrgp}. The predictions of the error from Section \ref{sec:GPRprediction} were obtained with the function \texttt{predict} from the Statistics and Machine Learning Toolbox. The step size $\nu$ was taken as 0.001, which will be explained in Section \ref{sec:Comparison1}.

\subsection{Application of Online Feedback Optimization with no mismatch}
\label{sec:ResultsNoMismatch}

\subsubsection{No mismatch in compressor characteristics}
\begin{table}[!tbp]
\centering
\caption{Parameters of efficiency with no mismatch}
\label{tbl:EfficiencyPolynomials}
\begin{tabular}{@{}lcccccc@{}}
\toprule
   & 1      & $m_i $    & $\Pi_i $    & $m_i\Pi_i$ & $m_i^2$ & $\Pi_i^2$   \\ \midrule
$\alpha^1_{\text{default}}$ & 0.5919 & -0.0021 & 0.2934  & 0.0030 & 0     & -0.1179 \\
$\alpha^2_{\text{default}}$ & 0.6383 & -0.0020 & 0.3220  & 0.0034 & 0     & -0.1260 \\
$\alpha^3_{\text{default}}$ & 0.6291 & -0.0023 & 0.3104  & 0.0032 & 0     & -0.1306 \\ \bottomrule
\end{tabular}
\end{table}

Table \ref{tbl:EfficiencyPolynomials} presents the coefficients in vector $\alpha^j=[\alpha_i]_{i=0,\ldots,5}$ from \eqref{eq:EfficiencyPoly} corresponding to the $j$-th compressor. The subscript $\mathsf{default}$ indicates the real values of the parameters. 

\begin{figure*}
     \centering
     \begin{subfigure}[b]{1\textwidth}
         \centering
\includegraphics[width=1\textwidth]{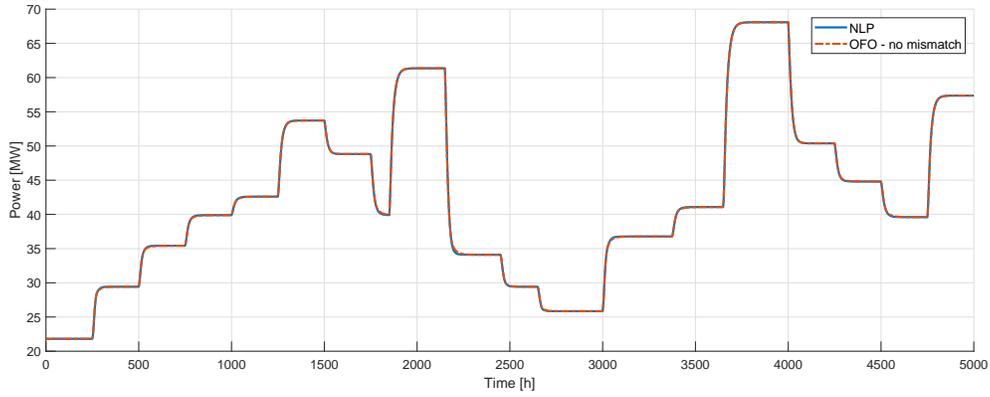}
\caption{Power consumption}
\label{fig:PowerAll_comparisonNoMismatch}
     \end{subfigure}
     \hfill
          \begin{subfigure}[b]{1\textwidth}
         \centering
\includegraphics[width=1\textwidth]{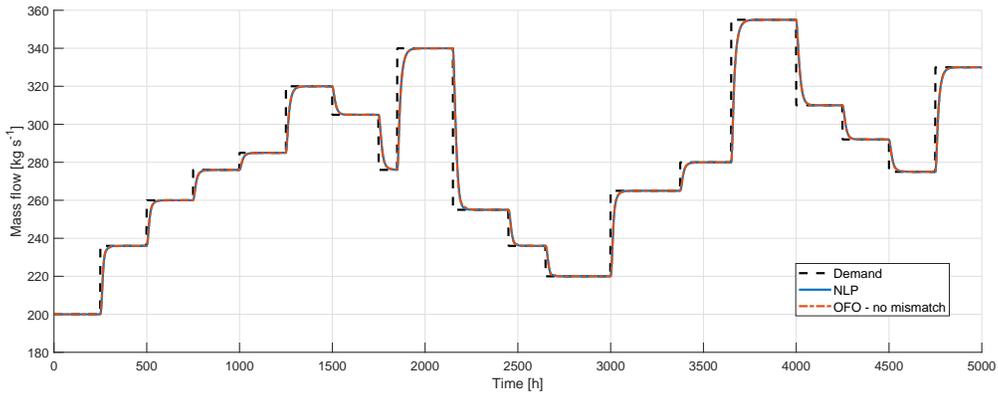}
\caption{Mass flow compared to the demand (thin black line)}
\label{fig:MassAll_comparisonNoMismatch}
     \end{subfigure}

        \caption{Comparison of the overall mass flow and power consumption from the compressor station if nonlinear optimization is used (solid blue line) and if \ac{OFO} with perfect knowledge is used (OFO, dash-dotted orange line)}
        \label{fig:Individuals_nomismatch}
\end{figure*}

\subsubsection{Comparison of approaches}
\label{sec:Comparison1}
In the first step, we show that \ac{OFO} is well-suited for load-sharing of a compressor station if there is no mismatch by comparing \ac{OFO} against the feedforward optimization based on nonlinear programming. We used this comparison to choose the tuning parameter $\nu$ from Eq. \eqref{eqn:Verena_sigma} in Section \ref{sec:FOcompressors}.

The overall power consumption of the station is similar in \ac{OFO} (dash-dotted line in Fig. \ref{fig:PowerAll_comparisonNoMismatch}) and the NLP (solid line in Fig. \ref{fig:PowerAll_comparisonNoMismatch}). The difference between the power consumption in both approaches is 0.2\%. The difference is due to transient behaviour of \ac{OFO} before reaching steady state, as explained in \cite{Online_Degner2021}. The transient behaviour in the mass flow and power will be further discussed in Section \ref{sec:Limitations}. Figure \ref{fig:MassAll_comparisonNoMismatch} shows also that the equality constraint \eqref{eq:DemandConstraint} was satisfied in both the NLP and OFO approach because both the blue and the orange lines follow the demand (black). Feedback optimization (OFO) reaches the same value of the mass flow as NLP in steady state, showing that OFO can handle equality constraints \eqref{eq:DemandConstraint} using the reformulation from \eqref{eq:DemandRelaxation}. Figure \ref{fig:MassAll_comparisonNoMismatch} also confirms theoretical properties from \cite{Non_Haeberle2020} that \ac{OFO} converges to the optimal solution, provided a steady-state input-output relationship is used. 

\begin{figure*}
     \centering
     \begin{subfigure}[b]{1\textwidth}
         \centering
         \includegraphics[width=\textwidth]{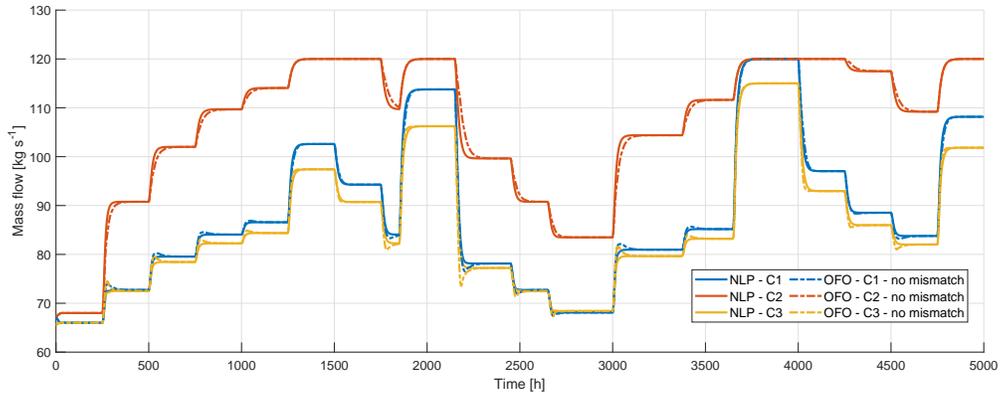}
         \caption{Individual mass flows}
         \label{fig:MassSep_comparisonNoMismatch}
     \end{subfigure}
          \hfill
               \begin{subfigure}[b]{1\textwidth}
         \centering
         \includegraphics[width=\textwidth]{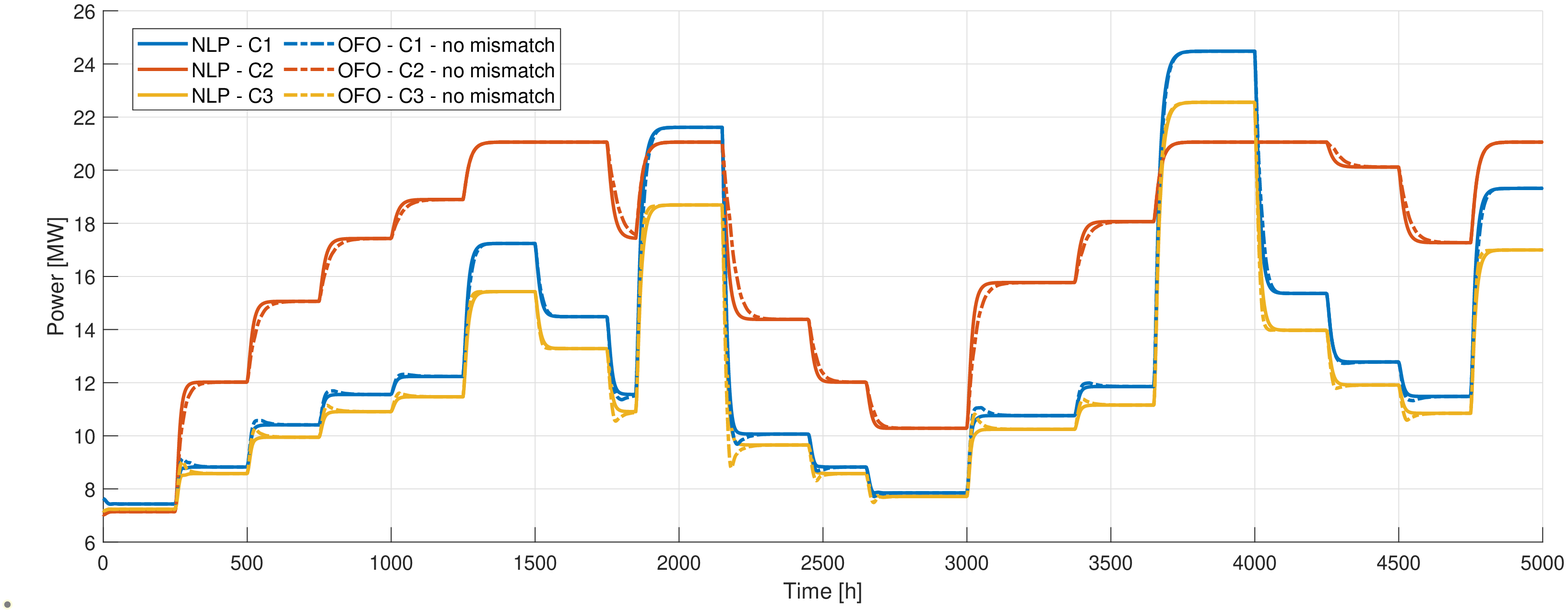}
         \caption{Individual power consumption}
         \label{fig:PowerSep_comparisonNoMismatch}
     \end{subfigure}
        \caption{Individual mass flow and power consumption for each compressor  if nonlinear optimization is used (NLP, solid line) and if \ac{OFO} with perfect knowledge is used (OFO, dash-dotted line)}
        \label{fig:IndividualsNoMismatch}
\end{figure*}

The \ac{OFO} controller operates each compressor at its optimal steady state as shown in Figs. \ref{fig:MassSep_comparisonNoMismatch} and \ref{fig:PowerSep_comparisonNoMismatch}. Both the mass flow through each compressor (Fig. \ref{fig:MassSep_comparisonNoMismatch}) and the power consumption (Fig. \ref{fig:PowerSep_comparisonNoMismatch}) achieve the same steady state in \ac{OFO} (dash-dotted lines) and nonlinear optimization (solid lines). The results from Fig. \ref{fig:IndividualsNoMismatch} together with the satisfaction of demand shown in Fig. \ref{fig:MassAll_comparisonNoMismatch} confirm that OFO is a suitable approach for operating compressor stations without explicitly solving full nonlinear optimization problems.

\subsection{Application of Online Feedback Optimization with mismatch}

\subsubsection{Mismatch in compressor characteristics}
\begin{figure}
     \centering
     \begin{subfigure}[b]{0.5\textwidth}
         \centering
         \includegraphics[width=\textwidth]{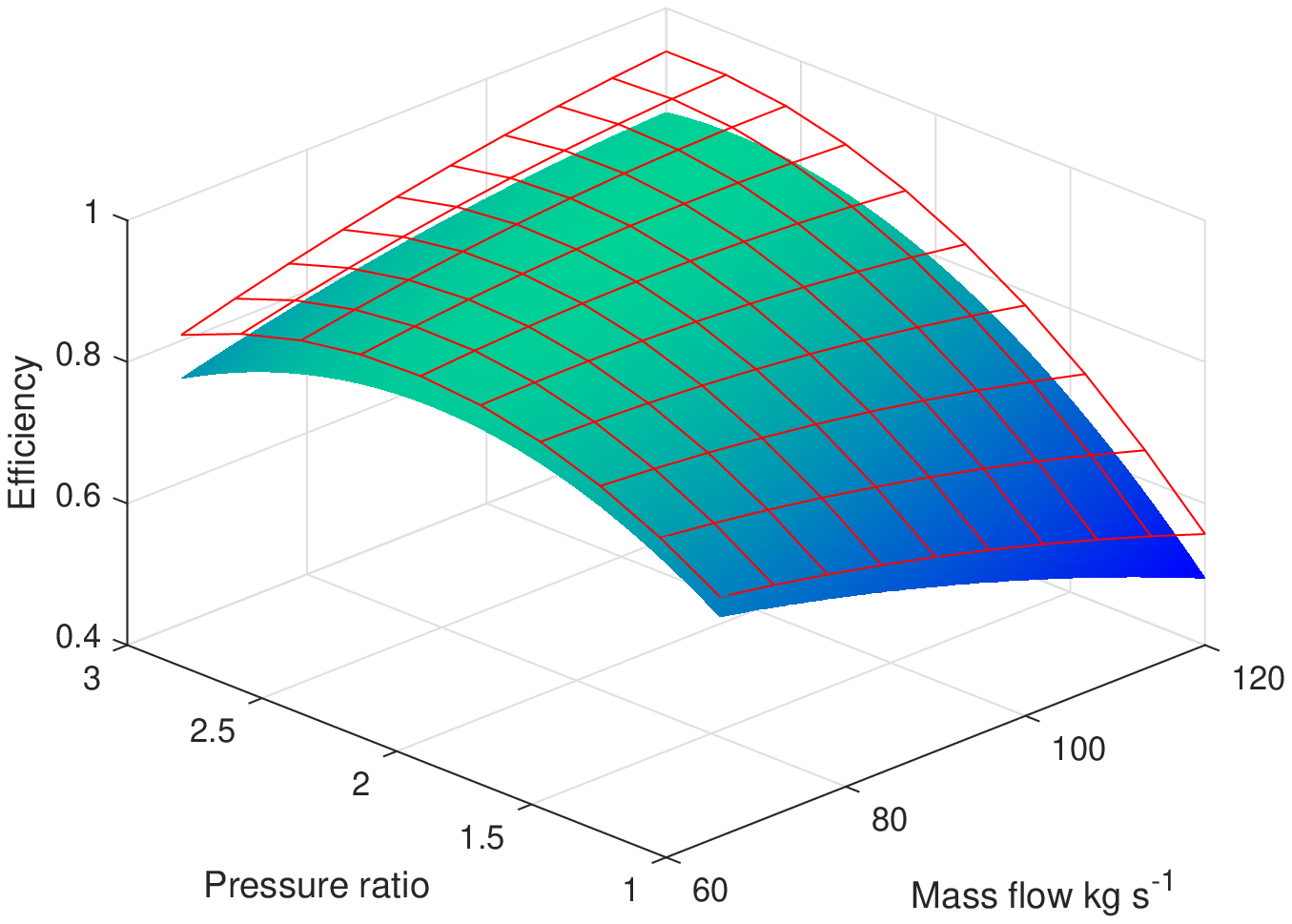}
         \caption{Mismatch for Compressor 1}
         \label{fig:MismatchC1}
     \end{subfigure}
     \hfill
     \begin{subfigure}[b]{0.5\textwidth}
         \centering
         \includegraphics[width=\textwidth]{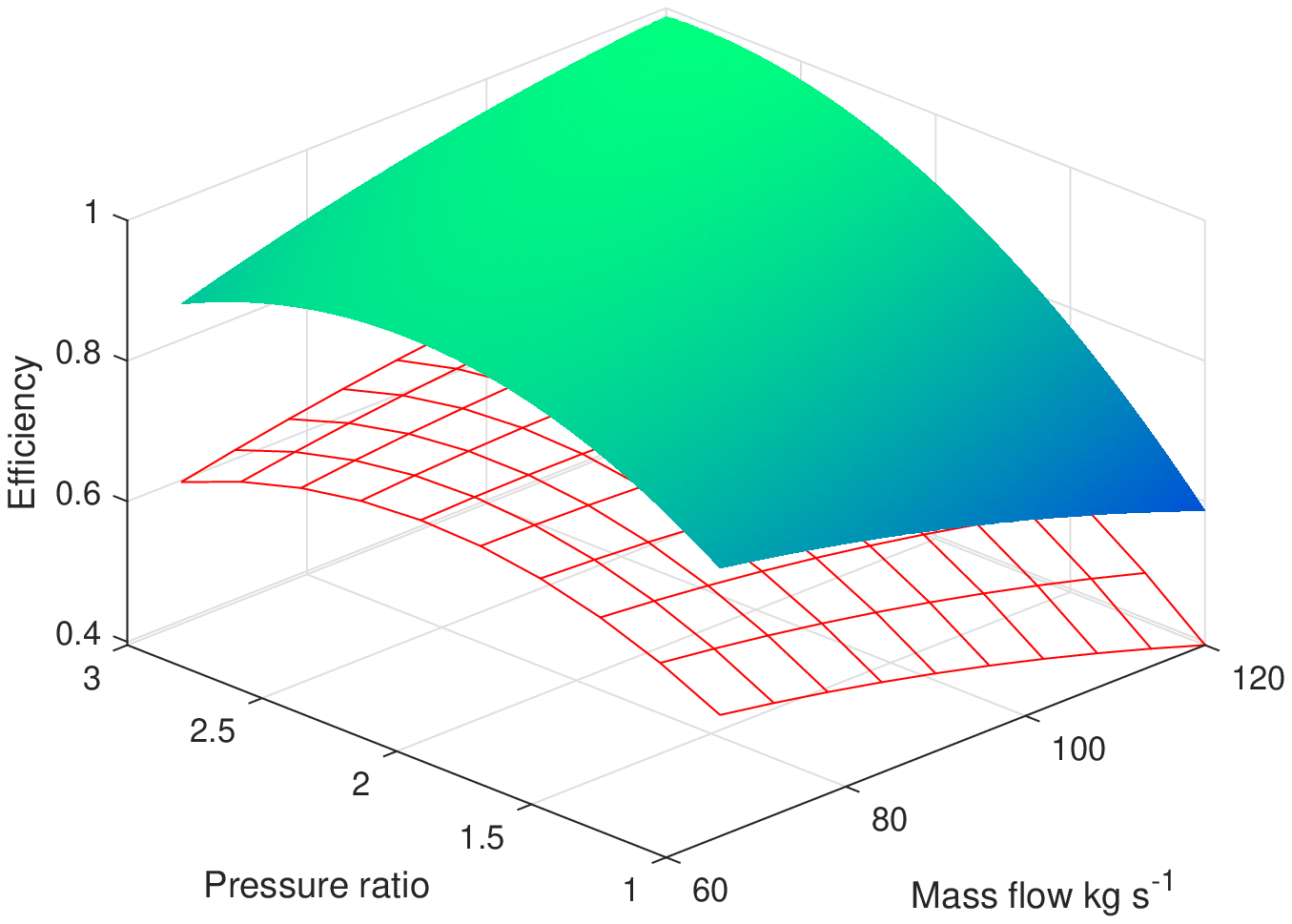}
         \caption{Mismatch for Compressor 2}
         \label{fig:MismatchC2}
     \end{subfigure}
     \hfill
     \begin{subfigure}[b]{0.5\textwidth}
         \centering
         \includegraphics[width=\textwidth]{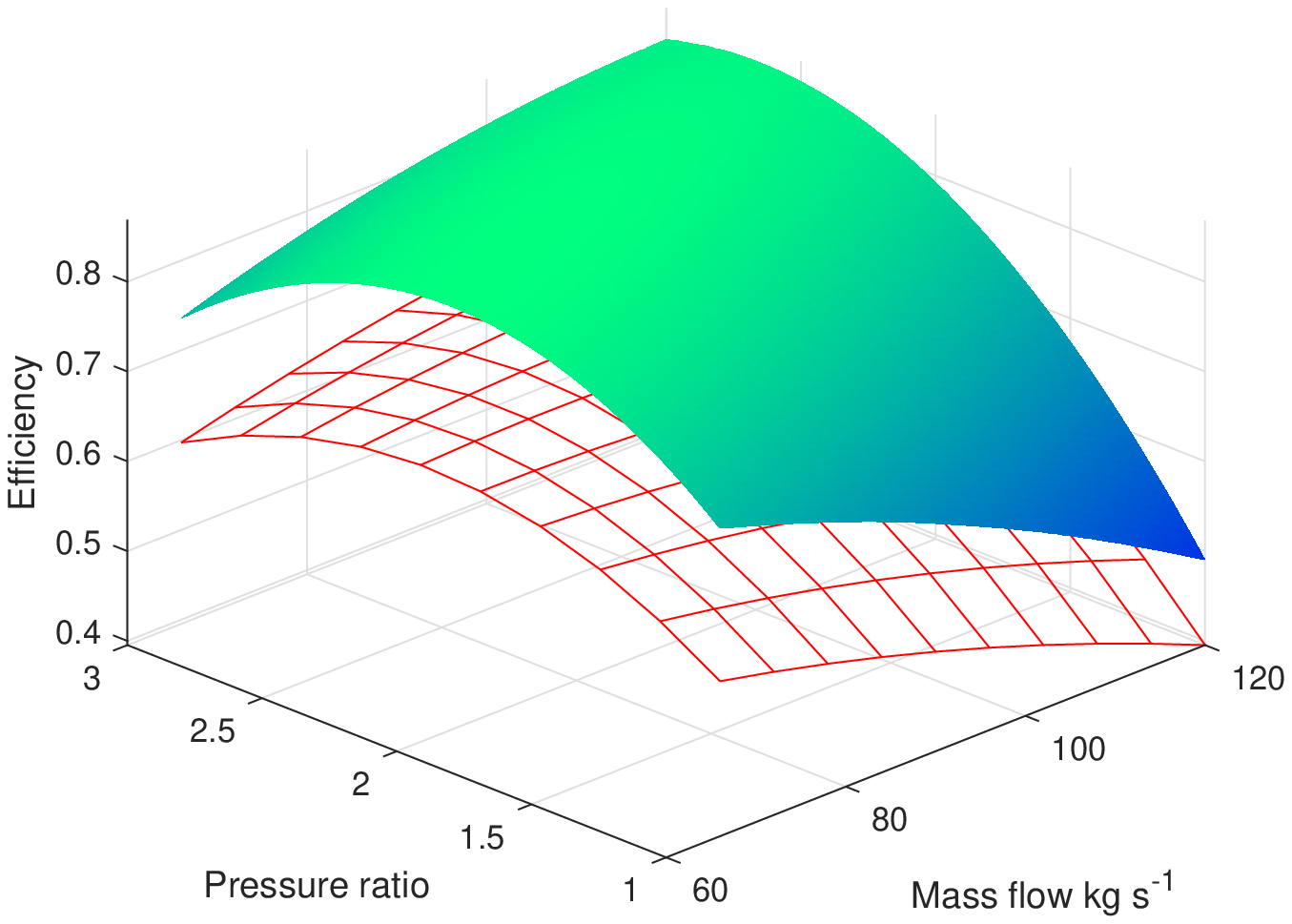}
         \caption{Mismatch for Compressor 3}
         \label{fig:MismatchC3}
     \end{subfigure}
        \caption{Efficiency mismatch, with multicoloured surface representing real compressor and red mesh representing the characteristics used in \ac{OFO}}
        \label{fig:Mismatch}
\end{figure}

The results from Section \ref{sec:ResultsNoMismatch} are now extended to show the impact of mismatch on the performance of OFO. The mismatch was introduced as follows
\begin{equation}
\begin{aligned}
\alpha^1_{\text{mismatch}}=&{}0.95\alpha^3_{\text{default}}\\
\alpha^2_{\text{mismatch}}=&{}0.8\alpha^1_{\text{default}}\\
\alpha^3_{\text{mismatch}}=&{}0.8\alpha^1_{\text{default}}
\end{aligned}
\label{eq:Mismatch}
\end{equation}
where the subscript ${\mathsf{default}}$ indicates the value of the real compressor. For instance, the real coefficients for Compressor $1$ are denoted by $\alpha^1_{\text{default}}$ whereas the coefficients for Compressor 1 with introduced mismatch are given by multiplying the real coefficients of Compressor 3 by 0.95.  Graphically, the mismatch is shown in Fig. \ref{fig:Mismatch}. The multicoloured surfaces represent the real compressors, whereas the transparent red meshes correspond to the efficiency characteristics used in \ac{OFO}. Compressor 1 (Fig. \ref{fig:MismatchC1}) has a lower real efficiency than assumed in OFO, whereas both Compressor 2 (Fig. \ref{fig:MismatchC2}) and 3 (Fig. \ref{fig:MismatchC3}) have a higher efficiency than assumed.

\subsubsection{Comparison between approaches}
\begin{figure*}
     \centering
     \begin{subfigure}[b]{1\textwidth}
         \centering
\includegraphics[width=1\textwidth]{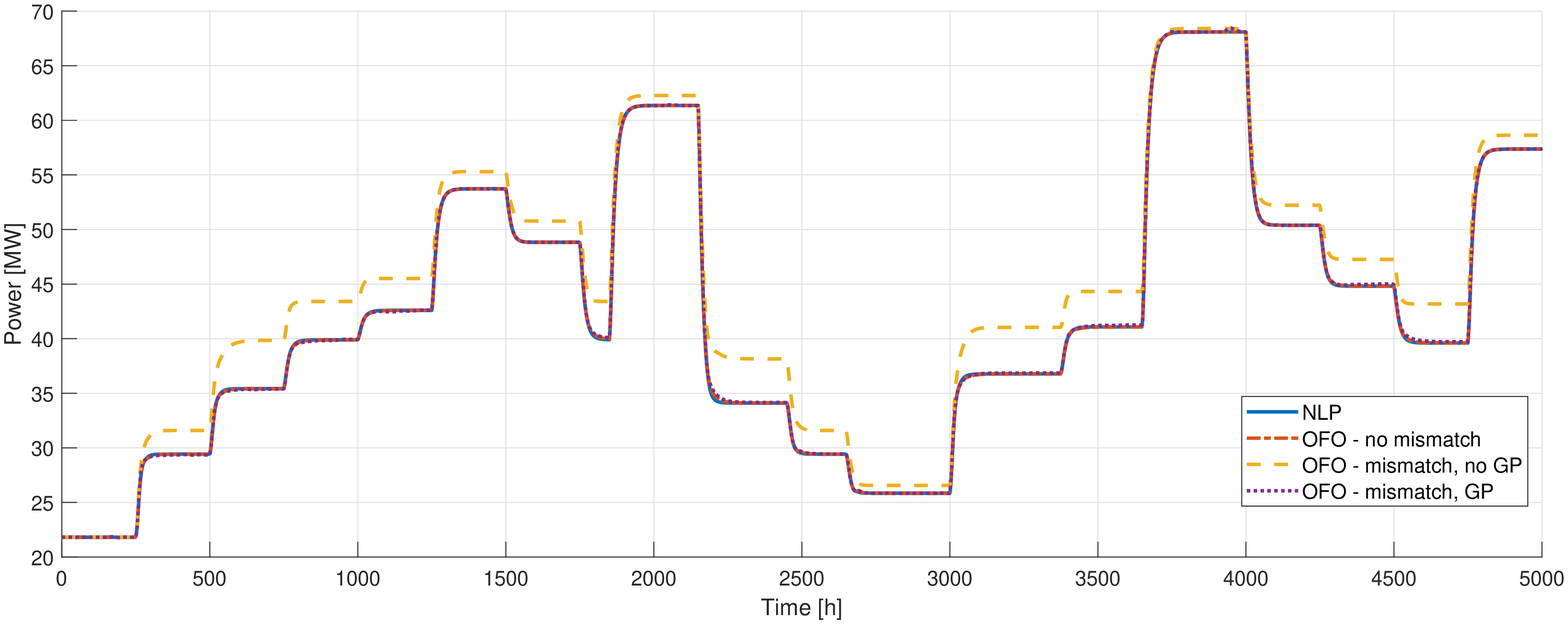}
\caption{Power consumption}
\label{fig:PowerAll_comparison}
     \end{subfigure}
     \hfill
          \begin{subfigure}[b]{1\textwidth}
         \centering
\includegraphics[width=1\textwidth]{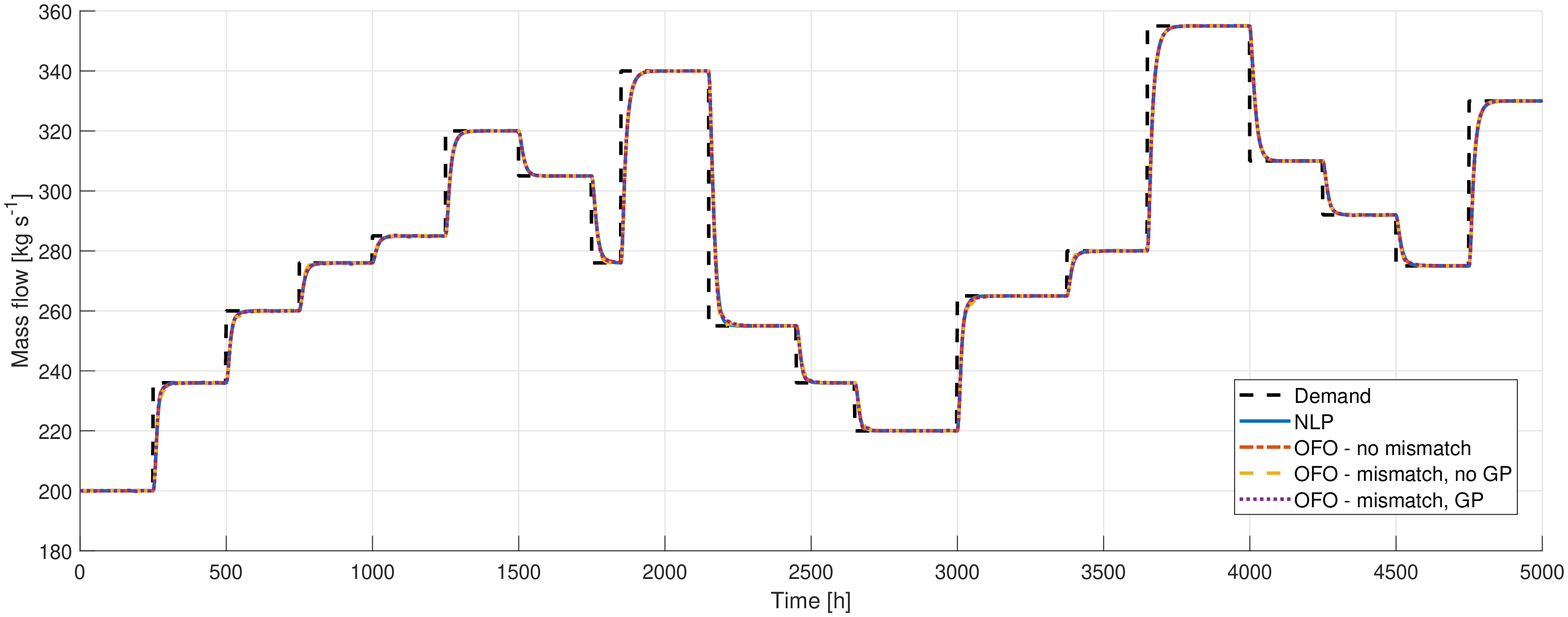}
\caption{Mass flow}
\label{fig:MassAll_comparison}
     \end{subfigure}

        \caption{Comparison of the overall mass flow and power consumption from the compressor station if nonlinear optimization is used (blue solid line), if \ac{OFO} with perfect knowledge is used (orange dash-dotted line), if \ac{OFO} without GP adaptation is used (yellow dashed line), and if \ac{OFO} with GP adaptation is used (purple dotted line)}
        \label{fig:Individuals_nomismatchAll}
\end{figure*}

\begin{figure*}
     \centering
     \begin{subfigure}[b]{1\textwidth}
         \centering
         \includegraphics[width=\textwidth]{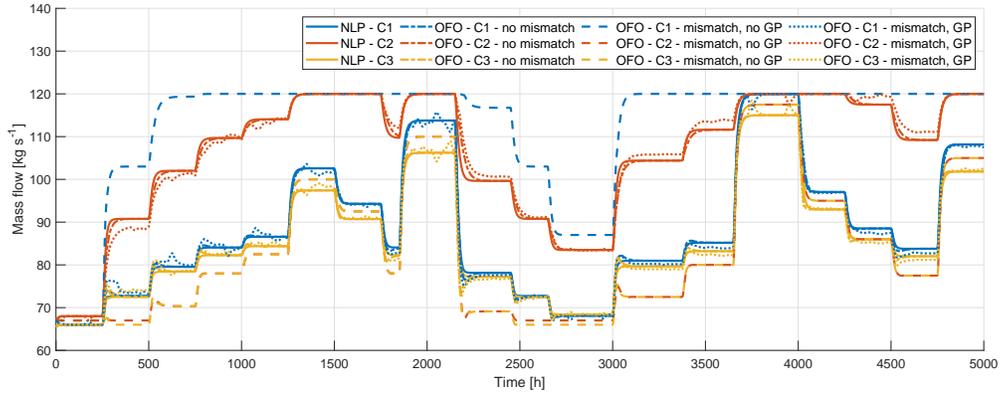}
         \caption{Individual mass flows}
         \label{fig:MassSep_comparison}
     \end{subfigure}
          \hfill
               \begin{subfigure}[b]{1\textwidth}
         \centering
         \includegraphics[width=\textwidth]{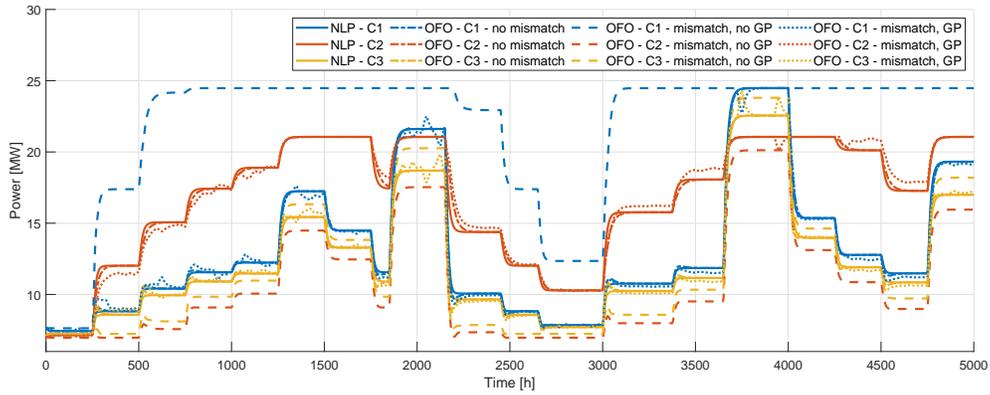}
         \caption{Individual power consumption}
         \label{fig:PowerSep_comparison}
     \end{subfigure}
        \caption{Individual mass flow and power consumption for each compressor  if nonlinear optimization is used (NLP, solid line), if \ac{OFO} with perfect knowledge is used (OFO, dash-dotted line), if \ac{OFO} without GP adaptation is used (OFO, dashed line), and if \ac{OFO} with GP adaptation is used (OFO, dotted line)}
        \label{fig:Individuals}
\end{figure*}

The impact of the mismatch together with the proposed mitigation based on Gaussian process regression is shown in Fig. \ref{fig:Individuals_nomismatchAll} and \ref{fig:Individuals}.

Figure \ref{fig:PowerAll_comparison} shows the overall power required to run the compressor station. The purple dotted line corresponding to \ac{OFO} with GP adaptation follows closely the benchmark results obtained from nonlinear optimization (solid blue line). The dashed orange line corresponding to \ac{OFO} without GP adaptation shows a higher power consumption of the station, which is due to the plant-model mismatch. Integrating the power consumption over time shows a 5\% increase in power consumption if no model adaptation was used, compared to 0.8\% increase if the proposed adaptation with Gaussian process regression was used. The increase is due to the learning period of Gaussian processes analysed in Section \ref{sec:MismatchLearning}.

Figure \ref{fig:MassAll_comparison} shows how the demand constraint from \eqref{eq:DemandConstraint} was satisfied by all four approaches. All three approaches based on \ac{OFO} satisfied the demand constraint in the same way as the benchmark solution based on nonlinear optimization. This proves that the reformulation of the equality constraint in \eqref{eq:DemandRelaxation} is sufficient for satisfaction of the equality constraint \eqref{eq:DemandConstraint}. Satisfaction of the equality constraint regardless of the mismatch proves that OFO can satisfy constraints without a priori solving nonlinear constrained optimization problems.

The difference between \ac{OFO} and the benchmark NLP solution can also be seen when looking at the flows and power consumption of the individual compressors, shown in Fig. \ref{fig:Individuals}. If there is no model adaptation, \ac{OFO} assigns loads that differ from the benchmark solution (NLP, solid lines). This is visible in particular for Compressor 1, shown in blue in Fig. \ref{fig:MassSep_comparison}. \ac{OFO} with no adaptation expects the efficiency of Compressor 1 to be higher (red mesh in Fig. \ref{fig:MismatchC1}) than the actual efficiency (multicoloured surface in Fig. \ref{fig:MismatchC1}). Therefore, the load assigned to Compressor 1 if there is no model adaptation is larger than the load assigned to Compressor 1 in the other two scenarios. The inability to assign a correct load to Compressor 1 led to increased power consumption of Compressor 1 as shown in Fig. \ref{fig:PowerSep_comparison} (dashed blue line).

\subsection{Mismatch learning}
\label{sec:MismatchLearning}

\subsubsection{Overall performance}
The adaptation of the model from Eq. \eqref{eq:EfficiencyPoly} was done every 25 h by re-tuning the Gaussian processes using past data. As a result, we obtained 200 Gaussian Processes over the course of 5000 hours. The demand is assumed to be known and the adaptation time step was chosen so that the efficiency estimation from Algorithm \ref{alg:AdaptationWithGPR} was conducted every time the demand changes.

\begin{figure*}
     \centering
         \includegraphics[width=\textwidth]{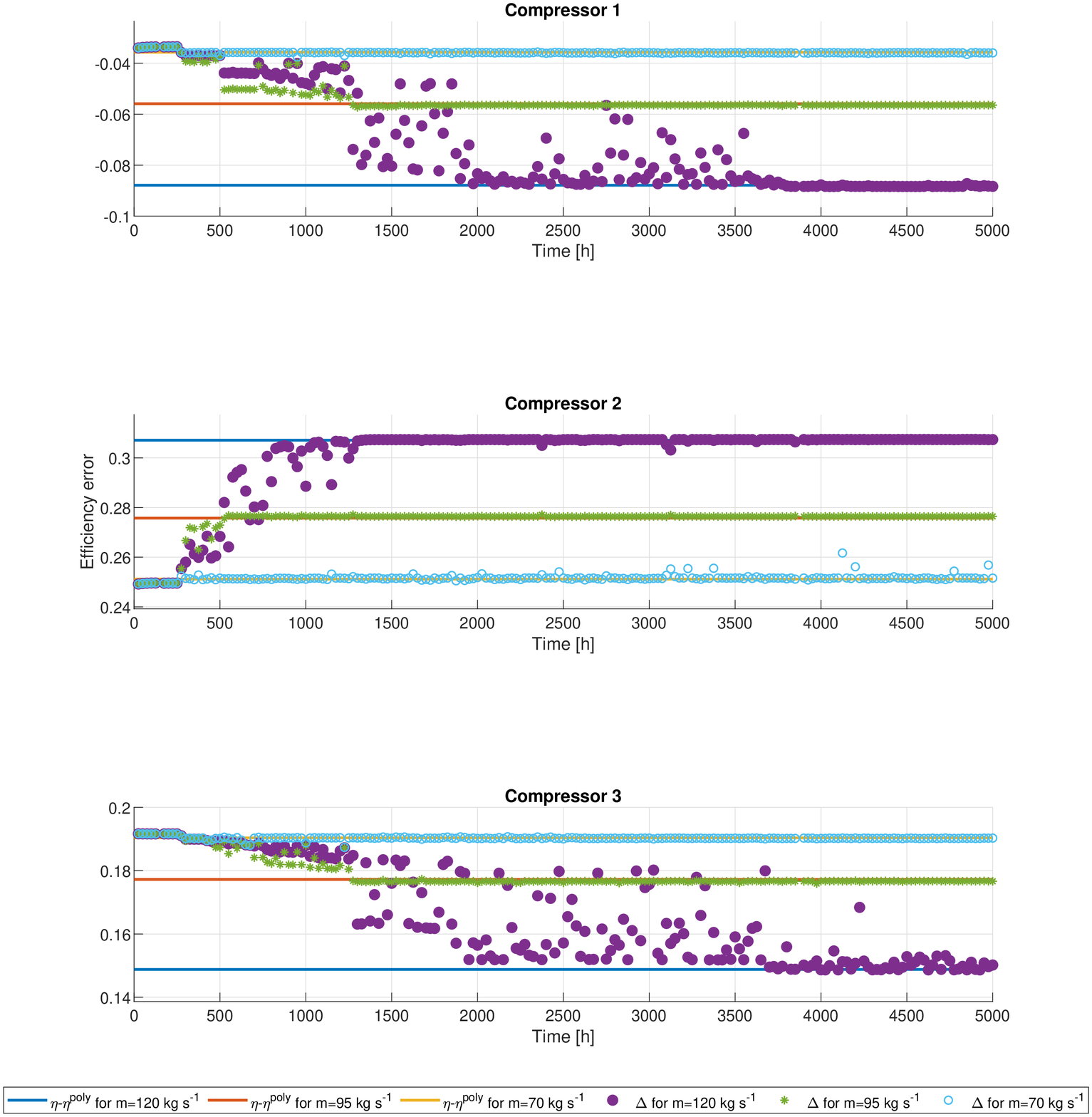}
        \caption{Learning process for Compressor 1 (top), Compressor 2 (middle), Compressor 3 (bottom)}
        \label{fig:AllLearning}
\end{figure*}

The learning progress is shown in Fig. \ref{fig:AllLearning}. To show the learning process, we plotted the error predicted by the GP obtained in a given time instant for three values of the mass flow: 70, 95, and 120 kg s$^{-1}$. The values of the mass flows were chosen to cover the whole range of compressor operating points. The solid lines in Fig. \ref{fig:AllLearning} show the error for a given mass flow (yellow - 70 kg s$^{-1}$, orange - 95 kg s$^{-1}$, blue - 120 kg s$^{-1}$) obtained from Eq. \eqref{eq:Error} as a difference between real efficiency and the polynomial approximation with mismatch for Compressor 1 (top), Compressor 2 (middle), Compressor 3 (bottom). The dots show the predicted error every 25h (light blue circles - 70 kg s$^{-1}$, green dots - 95 kg s$^{-1}$, purple - 120 kg s$^{-1}$).

The right-hand side of the three plots, near 5000 hours, shows that the GPs predict the true value of the error because the dots overlap with the solid lines. The overlap of the predicted values and the true error is particularly visible for Compressor 1 and Compressor 3 for all the mass flows. The reason for the overlap for these two compressors is that they both reach every value of the mass flow from 70 to 120 kg s$^{-1}$. At the same time, the largest mass flow reached by Compressor 3 is 115 kg s$^{-1}$ (yellow dotted lines in Fig. \ref{fig:MassSep_comparison} between 3500 and 4000 h). As a result, the mass flow of 120 kg s$^{-1}$ has never been included in the GPs for Compressor 3, resulting in larger differences between the true error and prediction (purple dots lie above the blue line in the bottom plot of Fig. \ref{fig:AllLearning}). However, these differences have little effect on the performance of \ac{OFO} shown in Fig. \ref{fig:MassSep_comparison} where the dotted lines representing the results of \ac{OFO} follow the solid lines of the NLP solution.

The fact that the GPs give better predictions for a given mass flow after the mass flow has been reached by a compressor is confirmed in the left-hand side of Fig. \ref{fig:AllLearning}. During the first 250 h, the error predicted by GPs for every mass flow differs from the true error for all three compressors (the dots are above the solid lines for Compressor 1 and 3, and below the solid line for Compressor 2). The difference appears because all the three compressors have initial mass flows different from either 70, 95, or 120 kg s$^{-1}$, as shown in Fig. \ref{fig:MassSep_comparison}. As soon as the set-points change and the three compressors reach mass flows larger than 70 kg s$^{-1}$, the new points are used for training the GPs. As a result, the GPs give accurate predictions for 70 kg s$^{-1}$ just after the first set-point change. In a similar way, the GPs give accurate prediction for 95 kg s$^{-1}$ for Compressor 1 and Compressor 3 after 1250 h, and for Compressor 2 after 500 h. Furthermore, as soon as the compressors reach 120 kg s$^{-1}$ (Compressor 1 after 3500 h and Compressor 2 after 1250), the GPs give accurate predictions for these mass flows. In particular, Compressor 2 works close to 70 kg s$^{-1}$ only at the very beginning, so that period is the only one with 70 kg s$^{-1}$ considered by the GPs. Thus, the points farther from 70 kg s$^{-1}$ have more impact on the learning process which results in the blue circles lying farther from the true error (compared to Compressor 1 and Compressor 2 for the mass flows 70 kg s$^{-1}$ at the end of the time period).

\subsubsection{Effects of mismatch}
To show the performance of model adaptation we ran the case study for 12 types of mismatch for every compressor. The chosen mismatch cases capture both parametric and structural mismatch. We assumed that the real characteristics of the compressors is either a quadratic polynomial of the form \eqref{eq:EfficiencyPoly} or a two-dimensional sinusoidal function:
\begin{equation}
    \eta_i = s^i_2\sin(0.02(m+s^i_3\Pi+s^i_1))
    \label{eq:SinApprox}
\end{equation}
parametrised by $\mathbf{s}^i=[s^i_1,s^i_2,s^i_3]$ for the $i$-th compressor. The values of $\mathbf{s}^i$ are given in Table \ref{tbl:SinModels}. Furthermore, the real characteristics have been affected by additive noise of magnitude $\pm 0.001$.

\begin{table}[tbp]
\caption{Parameters of the characteristics in the form \ref{eq:SinApprox}}
\label{tbl:SinModels}
\begin{tabular}{@{}llll@{}}
\toprule
      & $s^i_1$ & $s^i_2$ & $s^i_3$ \\ \midrule
$i=1$ & -7.294  & 0.8559  & -9.222  \\
$i=2$ & -11.15  & 0.966   & -7.511  \\
$i=3$ & -3.595  & 0.8584  & -10.47  \\ \bottomrule
\end{tabular}
\end{table}

The approximation $\hat{\eta}(\cdot,\cdot)$ of the characteristics used to evaluate the error in \eqref{eq:Error} was then taken as a quadratic polynomial, linear polynomial, and a constant value. These approximations were chosen based on how compressor efficiency is typically described in practice \citep{Experimental_Cortinovis2015}. The values of the parameters of the respective polynomials were obtained by setting the corresponding parameters in Table \ref{tbl:EfficiencyPolynomials} to zero.

The performance was evaluated using Mean Absolute Error (MAE) over $N$ time steps for the $i$-th compressor:
\begin{equation}
\text{MAE}_i = \frac{\sum_{j=1}^N|\epsilon^j_i-\varepsilon_i|}{N}    
\end{equation}
where $\epsilon^j_i$ was obtained from the Gaussian process in step $j$ and $\varepsilon_i$ is the actual error. The results for Compressor 1, Compressor 2, and Compressor 3 for the three mass flows are collected in Tables \ref{tbl:C1Assessment}, \ref{tbl:C2Assessment}, and \ref{tbl:C3Assessment}, respectively. We also show the estimation error obtained from the first Gaussian process, $\delta_{\text{init}}$ and the last Gaussian process, $\delta_{\text{fin}}$. Table \ref{tbl:C3Assessment} shows that regardless of the actual error value, MAE was smaller than 0.05, obtained for Compressor 3 for linear estimate of the polynomial characteristic with noise. In particular, the value of MAE for the largest error between the real characteristics and the estimate, obtained for Compressor 2 for constant estimate, was 0.007 (Table \ref{tbl:C2Assessment}). These results indicate that the Gaussian process provided a good approximation of the error over all time steps on average.

The learning process, similar to the one depicted in Fig. \ref{fig:AllLearning}, is particularly visible if the real characteristics is sinusoidal. For Compressor 2 with constant estimate, the error went from 0.243 in the first time step to 0.001 in the final time step (Table \ref{tbl:C2Assessment}). Therefore, the Gaussian process modelling the error is able to capture the structural mismatch as well as parametric mismatch.

\begin{table}[]
\caption{Error values for Compressor 1}
\label{tbl:C1Assessment}
\begin{tabular}{p{0.8cm}|p{1.5cm}lllllllll}
\multicolumn{1}{l|}{}             & $\hat{\eta}^{\text{poly}}_1$ & \multicolumn{3}{c}{Quadratic} & \multicolumn{3}{c}{Linear} & \multicolumn{3}{c}{Constant} \\ \hline
\multicolumn{1}{l|}{$\eta_1$}     & $m_1$ [kg s$^{-1}$]                   & 70       & 95       & 120     & 70      & 95      & 120    & 70       & 95      & 120     \\ \hline
\multirow{4}{*}{Poly}       & $\eta_1-\hat{\eta}^{\text{poly}}_1$                                     & -0.036   & -0.056   & -0.088  & -0.167  & -0.228  & -0.370 & 0.239    & 0.252   & 0.243   \\ \cline{2-11} 
                                  & MAE                                       & 0.000    & 0.004    & 0.016   & 0.002   & 0.007   & 0.043  & 0.001    & 0.002   & 0.003   \\
                                  & $\delta_{\text{init}}$                                & 0.002    & 0.022    & 0.054   & 0.001   & 0.062   & 0.204  & -0.001   & -0.014  & -0.005  \\
                                  & $\delta_{\text{fin}}$                               & 0.000    & -0.001   & -0.001  & 0.000   & 0.000   & -0.001 & 0.001    & 0.000   & -0.001  \\ \cline{2-11} 
\multirow{3}{*}{+ noise} & MAE                                       & 0.000    & 0.004    & 0.014   & 0.002   & 0.008   & 0.041  & 0.001    & 0.001   & 0.004   \\
                                  & $\delta_{\text{init}}$                                & 0.002    & 0.022    & 0.054   & 0.000   & 0.062   & 0.203  & -0.002   & -0.014  & -0.005  \\
                                  & $\delta_{\text{fin}}$                               & 0.000    & -0.001   & -0.001  & 0.000   & 0.000   & -0.001 & 0.001    & 0.000   & -0.001  \\ \hline \hline
\multirow{4}{*}{Sin}       & $\eta_1-\hat{\eta}^{\text{poly}}_1$                                      & -0.214   & -0.087   & -0.088  & -0.345  & -0.259  & -0.370 & 0.061    & 0.221   & 0.243   \\ \cline{2-11} 
                                  & MAE                                       & 0.003    & 0.009    & 0.020   & 0.002   & 0.009   & 0.011  & 0.003    & 0.011   & 0.021   \\
                                  & $\delta_{\text{init}}$                                & -0.027   & -0.155   & -0.154  & -0.028  & -0.114  & -0.004 & -0.030   & -0.191  & -0.213  \\
                                  & $\delta_{\text{fin}}$                               & -0.001   & 0.000    & 0.002   & 0.000   & 0.000   & 0.000  & -0.001   & -0.001  & 0.001   \\ \cline{2-11} 
\multirow{3}{*}{+ noise} & MAE                                       & 0.004    & 0.017    & 0.024   & 0.002   & 0.008   & 0.010  & 0.003    & 0.011   & 0.027   \\
                                  & $\delta_{\text{init}}$                                & -0.027   & -0.155   & -0.154  & -0.029  & -0.115  & -0.004 & -0.031   & -0.191  & -0.213  \\
                                  & $\delta_{\text{fin}}$                               & -0.001   & 0.000    & 0.002   & 0.000   & 0.000   & -0.002 & -0.001   & -0.001  & 0.001  
\end{tabular}
\end{table}

\begin{table}[]
\caption{Error values for Compressor 2}
\label{tbl:C2Assessment}
\begin{tabular}{p{0.8cm}|p{1.5cm}lllllllll}                         & $\hat{\eta}^{\text{poly}}_2$        & \multicolumn{3}{c}{Quadratic}                                             & \multicolumn{3}{c}{Linear}                                                & \multicolumn{3}{c}{Constant}                                              \\ \hline
$\eta_2$                 & $m_2$ [kg s$^{-1}$]                 & \multicolumn{1}{l}{70} & \multicolumn{1}{l}{95} & \multicolumn{1}{l}{120} & \multicolumn{1}{l}{70} & \multicolumn{1}{l}{95} & \multicolumn{1}{l}{120} & \multicolumn{1}{l}{70} & \multicolumn{1}{l}{95} & \multicolumn{1}{l}{120} \\ \hline
\multirow{4}{*}{Poly}    & $\eta_2-\hat{\eta}^{\text{poly}}_2$ & 0.251                  & 0.276                  & 0.307                   & 0.115                  & 0.089                  & 0.002                   & 0.454                  & 0.489                  & 0.513                   \\ \cline{2-11} 
                         & MAE                                 & 0.001                  & 0.002                  & 0.007                   & 0.003                  & 0.003                  & 0.017                   & 0.001                  & 0.003                  & 0.007                   \\
                         & $\delta_{\text{init}}$              & -0.002                 & -0.027                 & -0.058                  & -0.002                 & 0.024                  & 0.111                   & -0.003                 & -0.038                 & -0.062                  \\
                         & $\delta_{\text{fin}}$               & 0.000                  & 0.001                  & 0.000                   & -0.002                 & 0.000                  & -0.002                  & 0.000                  & 0.001                  & -0.001                  \\ \cline{2-11} 
\multirow{3}{*}{+ noise} & MAE                                 & 0.001                  & 0.002                  & 0.007                   & 0.002                  & 0.003                  & 0.018                   & 0.001                  & 0.003                  & 0.007                   \\
                         & $\delta_{\text{init}}$              & -0.002                 & -0.026                 & -0.058                  & -0.002                 & 0.024                  & 0.110                   & -0.004                 & -0.038                 & -0.063                  \\
                         & $\delta_{\text{fin}}$               & 0.001                  & 0.000                  & 0.000                   & 0.000                  & 0.000                  & -0.001                  & 0.001                  & 0.001                  & -0.001                  \\ \hline \hline
\multirow{4}{*}{Sin}     & $\eta_2-\hat{\eta}^{\text{poly}}_2$ & 0.070                  & 0.248                  & 0.276                   & -0.066                 & 0.061                  & -0.029                  & 0.273                  & 0.462                  & 0.482                   \\ \cline{2-11} 
                         & MAE                                 & 0.005                  & 0.012                  & 0.023                   & 0.002                  & 0.009                  & 0.011                   & 0.006                  & 0.013                  & 0.019                   \\
                         & $\delta_{\text{init}}$              & -0.032                 & -0.210                 & -0.238                  & -0.032                 & -0.160                 & -0.069                  & -0.033                 & -0.222                 & -0.242                  \\
                         & $\delta_{\text{fin}}$               & -0.002                 & -0.001                 & 0.002                   & 0.000                  & 0.000                  & 0.000                   & -0.002                 & -0.001                 & 0.001                   \\ \cline{2-11} 
\multirow{3}{*}{+ noise} & MAE                                 & 0.006                  & 0.011                  & 0.022                   & 0.002                  & 0.009                  & 0.007                   & 0.004                  & 0.014                  & 0.026                   \\
                         & $\delta_{\text{init}}$              & -0.032                 & -0.210                 & -0.238                  & -0.033                 & -0.160                 & -0.070                  & -0.034                 & -0.222                 & -0.243                  \\
                         & $\delta_{\text{fin}}$               & -0.002                 & -0.001                 & 0.002                   & 0.001                  & 0.000                  & -0.001                  & -0.001                 & -0.001                 & 0.001                  
\end{tabular}
\end{table}

\begin{table}[]
\caption{Error values for Compressor 3}
\label{tbl:C3Assessment}
\begin{tabular}{p{0.8cm}|p{1.5cm}lllllllll}    
                         & $\hat{\eta}^{\text{poly}}_3$        & \multicolumn{3}{c}{Quadratic}                                             & \multicolumn{3}{c}{Linear}                                                & \multicolumn{3}{c}{Constant}                                              \\ \hline
$\eta_3$                 & $m_3$ [kg s$^{-1}$]                 & \multicolumn{1}{l}{70} & \multicolumn{1}{l}{95} & \multicolumn{1}{l}{120} & \multicolumn{1}{l}{70} & \multicolumn{1}{l}{95} & \multicolumn{1}{l}{120} & \multicolumn{1}{l}{70} & \multicolumn{1}{l}{95} & \multicolumn{1}{l}{120} \\ \hline
\multirow{4}{*}{Poly}    & $\eta_3-\hat{\eta}^{\text{poly}}_3$ & 0.190                  & 0.177                  & 0.149                   & 0.054                  & -0.010                 & -0.156                  & 0.393                  & 0.391                  & 0.355                   \\ \cline{2-11} 
                         & MAE                                 & 0.000                  & 0.003                  & 0.019                   & 0.001                  & 0.009                  & 0.048                   & 0.000                  & 0.001                  & 0.019                   \\
                         & $\delta_{\text{init}}$              & 0.001                  & 0.014                  & 0.043                   & 0.003                  & 0.067                  & 0.213                   & 0.000                  & 0.002                  & 0.038                   \\
                         & $\delta_{\text{fin}}$               & 0.000                  & -0.001                 & 0.000                   & -0.001                 & 0.000                  & -0.003                  & 0.000                  & 0.000                  & -0.001                  \\ \cline{2-11} 
\multirow{3}{*}{+ noise} & MAE                                 & 0.000                  & 0.003                  & 0.017                   & 0.001                  & 0.008                  & 0.050                   & 0.000                  & 0.001                  & 0.018                   \\
                         & $\delta_{\text{init}}$              & 0.002                  & 0.015                  & 0.043                   & 0.003                  & 0.067                  & 0.213                   & 0.000                  & 0.003                  & 0.039                   \\
                         & $\delta_{\text{fin}}$               & 0.000                  & 0.000                  & -0.001                  & 0.000                  & 0.000                  & -0.001                  & 0.000                  & 0.000                  & -0.001                  \\ \hline \hline
\multirow{4}{*}{Sin}     & $\eta_3-\hat{\eta}^{\text{poly}}_3$ & 0.006                  & 0.146                  & 0.173                   & -0.130                 & -0.041                 & -0.132                  & 0.209                  & 0.359                  & 0.379                   \\ \cline{2-11} 
                         & MAE                                 & 0.007                  & 0.026                  & 0.040                   & 0.003                  & 0.010                  & 0.013                   & 0.008                  & 0.032                  & 0.042                   \\
                         & $\delta_{\text{init}}$              & -0.039                 & -0.179                 & -0.206                  & -0.037                 & -0.127                 & -0.036                  & -0.040                 & -0.191                 & -0.210                  \\
                         & $\delta_{\text{fin}}$               & -0.002                 & -0.001                 & 0.001                   & 0.000                  & 0.000                  & 0.000                   & -0.001                 & -0.001                 & -0.006                  \\ \cline{2-11} 
\multirow{3}{*}{+ noise} & MAE                                 & 0.004                  & 0.016                  & 0.035                   & 0.006                  & 0.020                  & 0.007                   & 0.004                  & 0.014                  & 0.036                   \\
                         & $\delta_{\text{init}}$              & -0.039                 & -0.179                 & -0.205                  & -0.037                 & -0.127                 & -0.035                  & -0.040                 & -0.190                 & -0.210                  \\
                         & $\delta_{\text{fin}}$               & -0.002                 & -0.001                 & 0.002                   & 0.000                  & 0.000                  & -0.001                  & -0.001                 & 0.000                  & 0.001                  
\end{tabular}
\end{table}

Finally we show how online feedback optimization handles the mismatch by assessing how well the demand was satisfied. The demand satisfaction was assessed using the Mean Absolute Error:
\begin{equation}
\text{MAE}_{\text{demand}} = \frac{\sum_{j=1}^N|\sum_{i=1}^3 m_{i,j}-M_j|}{N}    
\end{equation}
where $N$ is the number of time steps. If there is no mismatch, OFO satisfies the demand equally well as nonlinear optimization, to within  $\pm 0.9$ \% which corresponds to $\pm 3.15$  kg s$^{-1}$. For all the mismatch cases the demand was satisfied to within $\pm 1.1$\% of the actual demand, corresponding to $\pm 4$ kg s$^{-1}$ of the overall mass flow. The values obtained for all the types of mismatch are collected in Table \ref{tbl:MAEDemand}. 

\begin{table}[]
\caption{The values of MAE$_{\text{demand}}$ for all the types of mismatch, in kg s$^{-1}$. If there is no mismatch, the value is 3.15 kg s$^{-1}$}
\label{tbl:MAEDemand}
\begin{tabular}{@{}llll@{}}
\toprule
           & Quadratic & Linear & Constant \\ \midrule
Poly       & 3.05      & 3.11   & 2.90     \\
+noise & 2.94      & 3.80   & 3.21     \\
Sin        & 2.95      & 3.10   & 2.95     \\
+noise  & 2.91      & 3.32   & 2.86     \\ \bottomrule
\end{tabular}
\end{table}

\subsection{Discussion and directions for future work}
\label{sec:Limitations}
\subsubsection{Gaussian process regression}
Gaussian process regression is used in this paper to approximate unknown compressor characteristics independently from the way of controlling the station. Focusing on the approximation of compressor characteristics allows improving model fidelity which in turn can lead to increased usefulness of the model outside controller design. For instance, improved model fidelity can be used for economic optimization considering environmental impact of the operation of the station, as indicated by \cite{Techno_Kashani2014}. \cite{Application_Ahmed2021} have shown that Gaussian process regression is a useful tool for approximating compressor characteristics. They explored the non-parametric nature of Gaussian processes to approximate the error between the real characteristics and the assumed model without prior knowledge about the error.

Adaptation with GPs is attractive as it performs better compared to polynomials if fewer data points are available \citep{Data_Korkmaz2022}. In particular, the results from Section \ref{sec:MismatchLearning} indicate that GPs are able to accurately predict error values for set-points that have not yet been observed. Since compressors may have limited number of different operating points, it is useful to have tools that rely on few measured data-points. 

A possible limitation of the approach based on Gaussian process is the computational complexity of fitting Gaussian processes to a growing dataset. In the current paper, the GP regression was done using the whole history. Using the whole history allowed obtaining accurate models at the expense of increased computational effort \cite{Gaussian_Rasmussen2006}. There is potential for reducing the computational effort by reducing the number of points used for regression \citep{Application_Ahmed2021}. As shown by \cite{Data_Korkmaz2022}, the complexity can be reduced by using systematic subsampling strategies. The influence of the reduced dataset on the performance of feedback optimization remains a topic for future work. 

In the future, we would also like to exploit the quantification of uncertainty inherent to Gaussian processes to enable robustification of the online optimization against possible constraint violation.

\subsubsection{Online Feedback Optimization}
The paper presents an application of Online Feedback Optimization to optimal operation of a compressor station. We show that OFO achieves the same steady state performance as classic nonlinear optimization without explicitly solving the nonlinear optimization problem. The results from Section \ref{sec:Results} confirm that Online Feedback Optimization works well with Gaussian process regression. 

The simplicity of OFO is further accentuated by a single parameter that requires tuning, $\nu$ in  \eqref{eqn:Verena_feedback}. The work from \cite{Timescale_Hauswirth2021} provides initial insights into the choice of $\nu$. In the current work, $\nu$ has been tuned to ensure good performance if there is no plant-model mismatch. The analysis of tuning of OFO including the additional dynamic behaviour from Gaussian process regression remains an open question. 

The block diagram from Fig. \ref{fig:BlockDiagram} emphasizes that \ac{OFO} is independent from the model adaptation based on Gaussian processes. The only connection between the adapted model and \ac{OFO} is in the estimation of power consumption providing the derivatives evaluated at the current output. At the same time, as shown in Section \ref{sec:MismatchLearning}, increasing the fidelity of the model and of the derivatives improves the performance of \ac{OFO}. The independence of \ac{OFO} from the model has already been explored by \cite{Adaptive_Picallo2021} who presented a model-free variant of \ac{OFO} based on direct estimation of derivatives. However, theoretical analysis of robustness of \ac{OFO} to the fidelity of derivatives remains an open question.

\section{Conclusions}
\label{sec:Conclusions}
Modelling errors and changes in system characteristics result  in plant-model mismatch and affects the performance of the controllers. Existing approaches for mitigating plant-model mismatch in industrial systems focus on improving model fidelity.  This paper addresses the mismatch by improving model fidelity using Gaussian process regression in an online controller based on Online Feedback Optimization (OFO). The novelty of the current paper consists in explicit improvement of the model of compressors used in \ac{OFO} by performing online Gaussian process regression to mitigate plant-model mismatch. The proposed approach was applied in a compressor station with three compressors. The paper shows that:
\begin{itemize}
    \item Online Feedback Optimization leads to the same solution as nonlinear optimization, without explicitly solving the nonlinear optimization problem;
    \item Gaussian process regression enables learning the mismatch between the plant and model online, both for parametric and structural mismatch;
    \item Online Feedback Optimization with Gaussian process regression mitigates the mismatch while satisfying the demand.
\end{itemize}

If there is no mismatch, the approach based on \ac{OFO} reaches the same solution for the three compressors as nonlinear optimization. A plant-model mismatch results in 5\% increase in power consumption in the compressor station, compared to the solution of nonlinear optimization and the \ac{OFO} solution obtained when there is no mismatch. The Gaussian process regression is then used to learn characteristics of individual compressors online. The combination of the new model adaptation approach based on Gaussian process regression with Online Feedback Optimization mitigates the increase in power consumption from 5\% to 0.8\%. The proposed approach is also able to handle both parametric and structural mismatch, while satisfying the required demand.

In future work, we would like to explore the stochastic nature of Gaussian process regression, as well as analyse theoretical properties of Online Feedback Optimization with model adaptation.

\section*{Acknowledgement}

Financial support from ABB for the Autonomous Industrial Systems Laboratory is gratefully acknowledged.

\bibliographystyle{elsarticle-harv} 
\bibliography{biblio}

\begin{thebibliography}{33}
\expandafter\ifx\csname natexlab\endcsname\relax\def\natexlab#1{#1}\fi
\providecommand{\url}[1]{\texttt{#1}}
\providecommand{\href}[2]{#2}
\providecommand{\path}[1]{#1}
\providecommand{\DOIprefix}{doi:}
\providecommand{\ArXivprefix}{arXiv:}
\providecommand{\URLprefix}{URL: }
\providecommand{\Pubmedprefix}{pmid:}
\providecommand{\doi}[1]{\href{http://dx.doi.org/#1}{\path{#1}}}
\providecommand{\Pubmed}[1]{\href{pmid:#1}{\path{#1}}}
\providecommand{\bibinfo}[2]{#2}
\ifx\xfnm\relax \def\xfnm[#1]{\unskip,\space#1}\fi
\bibitem[{Ahmed et~al.(2022)Ahmed, Zagorowska, del Rio-Chanona and
  Mercang{\"o}z}]{Application_Ahmed2021}
\bibinfo{author}{Ahmed, A.}, \bibinfo{author}{Zagorowska, M.},
  \bibinfo{author}{del Rio-Chanona, E.A.}, \bibinfo{author}{Mercang{\"o}z, M.},
  \bibinfo{year}{2022}.
\newblock \bibinfo{title}{Application of {G}aussian processes to online
  approximation of compressor maps for load-sharing in a compressor station},
  in: \bibinfo{booktitle}{European Control Conference (ECC), 12-15 July,
  London, UK}.
\bibitem[{BSI(2014)}]{BS_BSI2014}
\bibinfo{author}{BSI}, \bibinfo{year}{2014}.
\newblock \bibinfo{title}{{BS EN} 12583:2014. Gas infrastructure. Compressor
  stations - Functional requirements}.
\newblock \bibinfo{type}{Standard}. {British Standards Institution}.
\bibitem[{Chu et~al.(2018)Chu, Dai, Lu, Ma and Wang}]{Improved_Chu2018}
\bibinfo{author}{Chu, F.}, \bibinfo{author}{Dai, B.}, \bibinfo{author}{Lu, N.},
  \bibinfo{author}{Ma, X.}, \bibinfo{author}{Wang, F.}, \bibinfo{year}{2018}.
\newblock \bibinfo{title}{Improved fast model migration method for centrifugal
  compressor based on {B}ayesian algorithm and {G}aussian process model}.
\newblock \bibinfo{journal}{Science China Technological Sciences}
  \bibinfo{volume}{61}, \bibinfo{pages}{1950--1958}.
\bibitem[{Cortinovis et~al.(2015)Cortinovis, Ferreau, Lewandowski and
  Mercang{\"o}z}]{Experimental_Cortinovis2015}
\bibinfo{author}{Cortinovis, A.}, \bibinfo{author}{Ferreau, H.J.},
  \bibinfo{author}{Lewandowski, D.}, \bibinfo{author}{Mercang{\"o}z, M.},
  \bibinfo{year}{2015}.
\newblock \bibinfo{title}{Experimental evaluation of {MPC}-based anti-surge and
  process control for electric driven centrifugal gas compressors}.
\newblock \bibinfo{journal}{Journal of Process Control} \bibinfo{volume}{34},
  \bibinfo{pages}{13--25}.
\bibitem[{Cortinovis et~al.(2016)Cortinovis, Mercang{\"o}z, Zovadelli,
  Pareschi, De~Marco and Bittanti}]{Online_Cortinovis2016}
\bibinfo{author}{Cortinovis, A.}, \bibinfo{author}{Mercang{\"o}z, M.},
  \bibinfo{author}{Zovadelli, M.}, \bibinfo{author}{Pareschi, D.},
  \bibinfo{author}{De~Marco, A.}, \bibinfo{author}{Bittanti, S.},
  \bibinfo{year}{2016}.
\newblock \bibinfo{title}{Online performance tracking and load sharing
  optimization for parallel operation of gas compressors}.
\newblock \bibinfo{journal}{Computers and Chemical Engineering}
  \bibinfo{volume}{88}, \bibinfo{pages}{145--156}.
\bibitem[{Degner(2021)}]{Online_Degner2021}
\bibinfo{author}{Degner, M.}, \bibinfo{year}{2021}.
\newblock \bibinfo{title}{Online {F}eedback {O}ptimization for {G}as
  {C}ompressors}.
\newblock \bibinfo{note}{{ETH Z}urich,
  \url{https://doi.org/10.3929/ethz-b-000502040}}.
\bibitem[{Egeland and Gravdahl(2002)}]{Modeling_Egeland2002a}
\bibinfo{author}{Egeland, O.}, \bibinfo{author}{Gravdahl, J.T.},
  \bibinfo{year}{2002}.
\newblock \bibinfo{title}{Modeling and simulation for automatic control}.
  volume~\bibinfo{volume}{76}.
\newblock \bibinfo{publisher}{Marine Cybernetics Trondheim, Norway}.
\bibitem[{Gentsch and King(2020)}]{Real_Gentsch2020}
\bibinfo{author}{Gentsch, M.}, \bibinfo{author}{King, R.},
  \bibinfo{year}{2020}.
\newblock \bibinfo{title}{Real-time estimation of a multi-stage centrifugal
  compressor performance map considering real-gas processes and flexible
  operation}.
\newblock \bibinfo{journal}{Journal of Process Control} \bibinfo{volume}{85},
  \bibinfo{pages}{227--243}.
\bibitem[{H{\"a}berle et~al.(2020)H{\"a}berle, Hauswirth, Ortmann, Bolognani
  and D{\"o}rfler}]{Non_Haeberle2020}
\bibinfo{author}{H{\"a}berle, V.}, \bibinfo{author}{Hauswirth, A.},
  \bibinfo{author}{Ortmann, L.}, \bibinfo{author}{Bolognani, S.},
  \bibinfo{author}{D{\"o}rfler, F.}, \bibinfo{year}{2020}.
\newblock \bibinfo{title}{Non-convex feedback optimization with input and
  output constraints}.
\newblock \bibinfo{journal}{IEEE Control Systems Letters} \bibinfo{volume}{5},
  \bibinfo{pages}{343--348}.
\bibitem[{Hauswirth et~al.(2021a)Hauswirth, Bolognani, Hug and
  D{\"o}rfler}]{Optimization_Hauswirth2021}
\bibinfo{author}{Hauswirth, A.}, \bibinfo{author}{Bolognani, S.},
  \bibinfo{author}{Hug, G.}, \bibinfo{author}{D{\"o}rfler, F.},
  \bibinfo{year}{2021}a.
\newblock \bibinfo{title}{Optimization algorithms as robust feedback
  controllers}.
\newblock \bibinfo{journal}{preprint arXiv:2103.11329} .
\bibitem[{Hauswirth et~al.(2021b)Hauswirth, Bolognani, Hug and
  Dorfler}]{Timescale_Hauswirth2021}
\bibinfo{author}{Hauswirth, A.}, \bibinfo{author}{Bolognani, S.},
  \bibinfo{author}{Hug, G.}, \bibinfo{author}{Dorfler, F.},
  \bibinfo{year}{2021}b.
\newblock \bibinfo{title}{Timescale separation in autonomous optimization}.
\newblock \bibinfo{journal}{{IEEE} Transactions on Automatic Control}
  \bibinfo{volume}{66}, \bibinfo{pages}{611--624}.
\bibitem[{Jenicek et~al.(1995)Jenicek, Kralik et~al.}]{Optimized_Jenicek1995}
\bibinfo{author}{Jenicek, T.}, \bibinfo{author}{Kralik, J.}, et~al.,
  \bibinfo{year}{1995}.
\newblock \bibinfo{title}{Optimized control of generalized compressor station},
  in: \bibinfo{booktitle}{{PSIG} Annual Meeting, 18-20 October, Albuquerque,
  New Mexico}, \bibinfo{organization}{Pipeline Simulation Interest Group}.
\bibitem[{Jung et~al.(2017)Jung, Lee, Park, Kim, Lee and
  Han}]{Improved_Jung2017}
\bibinfo{author}{Jung, J.}, \bibinfo{author}{Lee, W.J.}, \bibinfo{author}{Park,
  S.}, \bibinfo{author}{Kim, Y.}, \bibinfo{author}{Lee, C.J.},
  \bibinfo{author}{Han, C.}, \bibinfo{year}{2017}.
\newblock \bibinfo{title}{Improved control strategy for fixed-speed compressors
  in parallel system}.
\newblock \bibinfo{journal}{Journal of Process Control} \bibinfo{volume}{53},
  \bibinfo{pages}{57--69}.
\bibitem[{Kashani and Molaei(2014)}]{Techno_Kashani2014}
\bibinfo{author}{Kashani, A.H.A.}, \bibinfo{author}{Molaei, R.},
  \bibinfo{year}{2014}.
\newblock \bibinfo{title}{Techno-economical and environmental optimization of
  natural gas network operation}.
\newblock \bibinfo{journal}{Chemical Engineering Research and Design}
  \bibinfo{volume}{92}, \bibinfo{pages}{2106--2122}.
\bibitem[{Korkmaz and Mercangöz(2022)}]{Data_Korkmaz2022}
\bibinfo{author}{Korkmaz, B.S.}, \bibinfo{author}{Mercangöz, M.},
  \bibinfo{year}{2022}.
\newblock \bibinfo{title}{Data driven modelling of centrifugal compressor maps
  forcontrol and optimization applications}, in: \bibinfo{booktitle}{{ECC
  2022}}.
\bibitem[{Kumar and Cortinovis(2017)}]{Load_Kumar2017}
\bibinfo{author}{Kumar, S.}, \bibinfo{author}{Cortinovis, A.},
  \bibinfo{year}{2017}.
\newblock \bibinfo{title}{Load sharing optimization for parallel and serial
  compressor stations}, in: \bibinfo{booktitle}{2017 IEEE Conference on Control
  Technology and Applications ({CCTA})}, \bibinfo{organization}{IEEE}. pp.
  \bibinfo{pages}{499--504}.
\newblock \DOIprefix\doi{10.1109/CCTA.2017.8062511}.
\bibitem[{Liptak(2005)}]{Instrument_Liptak2005}
\bibinfo{author}{Liptak, B.G.}, \bibinfo{year}{2005}.
\newblock \bibinfo{title}{Instrument Engineers' Handbook, Volume Two: Process
  Control and Optimization}.
\newblock \bibinfo{publisher}{CRC Press}.
\bibitem[{Milosavljevic et~al.(2016)Milosavljevic, Cortinovis, Marchetti,
  Faulwasser, Mercang{\"o}z and Bonvin}]{Optimal_Milosavljevic2016}
\bibinfo{author}{Milosavljevic, P.}, \bibinfo{author}{Cortinovis, A.},
  \bibinfo{author}{Marchetti, A.G.}, \bibinfo{author}{Faulwasser, T.},
  \bibinfo{author}{Mercang{\"o}z, M.}, \bibinfo{author}{Bonvin, D.},
  \bibinfo{year}{2016}.
\newblock \bibinfo{title}{Optimal load sharing of parallel compressors via
  modifier adaptation}, in: \bibinfo{booktitle}{2016 IEEE Conference on Control
  Applications (CCA)}, pp. \bibinfo{pages}{1488--1493}.
\bibitem[{Milosavljevic et~al.(2020)Milosavljevic, Marchetti, Cortinovis,
  Faulwasser, Mercang{\"o}z and Bonvin}]{Real_Milosavljevic2020}
\bibinfo{author}{Milosavljevic, P.}, \bibinfo{author}{Marchetti, A.G.},
  \bibinfo{author}{Cortinovis, A.}, \bibinfo{author}{Faulwasser, T.},
  \bibinfo{author}{Mercang{\"o}z, M.}, \bibinfo{author}{Bonvin, D.},
  \bibinfo{year}{2020}.
\newblock \bibinfo{title}{Real-time optimization of load sharing for gas
  compressors in the presence of uncertainty}.
\newblock \bibinfo{journal}{Applied Energy} \bibinfo{volume}{272},
  \bibinfo{pages}{114883}.
\bibitem[{Mokhatab et~al.(2015)Mokhatab, Poe and Mak}]{Handbook_Mokhatab2015}
\bibinfo{author}{Mokhatab, S.}, \bibinfo{author}{Poe, W.A.},
  \bibinfo{author}{Mak, J.Y.}, \bibinfo{year}{2015}.
\newblock \bibinfo{title}{Handbook of Natural Gas Transmission and Processing}.
\newblock \bibinfo{publisher}{Elsevier Science}.
\bibitem[{N{\o}rsteb{\o}(2008)}]{Optimum_Noersteboe2008}
\bibinfo{author}{N{\o}rsteb{\o}, V.S.}, \bibinfo{year}{2008}.
\newblock \bibinfo{title}{Optimum Operation of Gas Export Systems}.
\newblock Ph.D. thesis.
\newblock \bibinfo{note}{{N}orwegian University of Science and Technology
  (NTNU)}.
\bibitem[{Ortmann et~al.(2020)Ortmann, Hauswirth, Caduff, Dörfler and
  Bolognani}]{Experimental_Ortmann2020}
\bibinfo{author}{Ortmann, L.}, \bibinfo{author}{Hauswirth, A.},
  \bibinfo{author}{Caduff, I.}, \bibinfo{author}{Dörfler, F.},
  \bibinfo{author}{Bolognani, S.}, \bibinfo{year}{2020}.
\newblock \bibinfo{title}{Experimental validation of feedback optimization in
  power distribution grids}.
\newblock \bibinfo{journal}{Electric Power Systems Research}
  \bibinfo{volume}{189}, \bibinfo{pages}{106782}.
\bibitem[{Osiadacz and Bell(1981)}]{Local_Osiadacz1981}
\bibinfo{author}{Osiadacz, A.}, \bibinfo{author}{Bell, D.J.},
  \bibinfo{year}{1981}.
\newblock \bibinfo{title}{A local optimization procedure for a gas-compressor
  station}.
\newblock \bibinfo{journal}{Optimal Control Applications and Methods}
  \bibinfo{volume}{{2}}, \bibinfo{pages}{239--250}.
\bibitem[{Paparella et~al.(2013)Paparella, Dom{\'\i}nguez, Cortinovis,
  Mercang{\"o}z, Pareschi and Bittanti}]{Load_Paparella2013}
\bibinfo{author}{Paparella, F.}, \bibinfo{author}{Dom{\'\i}nguez, L.},
  \bibinfo{author}{Cortinovis, A.}, \bibinfo{author}{Mercang{\"o}z, M.},
  \bibinfo{author}{Pareschi, D.}, \bibinfo{author}{Bittanti, S.},
  \bibinfo{year}{2013}.
\newblock \bibinfo{title}{Load sharing optimization of parallel compressors},
  in: \bibinfo{booktitle}{European Control Conference (ECC) 2013}, pp.
  \bibinfo{pages}{17--19}.
\newblock \DOIprefix\doi{10.23919/ECC.2013.6669697}.
\bibitem[{Picallo et~al.(2021)Picallo, Ortmann, Bolognani and
  D{\"o}rfler}]{Adaptive_Picallo2021}
\bibinfo{author}{Picallo, M.}, \bibinfo{author}{Ortmann, L.},
  \bibinfo{author}{Bolognani, S.}, \bibinfo{author}{D{\"o}rfler, F.},
  \bibinfo{year}{2021}.
\newblock \bibinfo{title}{Adaptive real-time grid operation via online feedback
  optimization with sensitivity estimation}.
\newblock \bibinfo{journal}{arXiv preprint arXiv:2110.00954} .
\bibitem[{Rasmussen(2003)}]{Gaussian_Rasmussen2003}
\bibinfo{author}{Rasmussen, C.E.}, \bibinfo{year}{2003}.
\newblock \bibinfo{title}{Gaussian processes in machine learning}, in:
  \bibinfo{booktitle}{Summer school on machine learning},
  \bibinfo{organization}{Springer}. pp. \bibinfo{pages}{63--71}.
\bibitem[{Rasmussen and Williams(2006)}]{Gaussian_Rasmussen2006}
\bibinfo{author}{Rasmussen, C.E.}, \bibinfo{author}{Williams, C.K.},
  \bibinfo{year}{2006}.
\newblock \bibinfo{title}{Gaussian processes for machine learning}.
  volume~\bibinfo{volume}{2}.
\newblock \bibinfo{publisher}{MIT Press Cambridge, MA}.
\bibitem[{Ren et~al.(2021)Ren, Zhou, Peng and Ou}]{Lei_2021}
\bibinfo{author}{Ren, L.}, \bibinfo{author}{Zhou, S.}, \bibinfo{author}{Peng,
  T.}, \bibinfo{author}{Ou, X.}, \bibinfo{year}{2021}.
\newblock \bibinfo{title}{A review of {CO}$_2$ emissions reduction technologies
  and low-carbon development in the iron and steel industry focusing on china}.
\newblock \bibinfo{journal}{Renewable and Sustainable Energy Reviews}
  \bibinfo{volume}{143}, \bibinfo{pages}{110846}.
\bibitem[{Riungu and Moses(2022)}]{Riungu_2022}
\bibinfo{author}{Riungu, E.K.}, \bibinfo{author}{Moses, P.M.},
  \bibinfo{year}{2022}.
\newblock \bibinfo{title}{Economic analysis and energy savings of variable
  speed drives in fans application — a case study {C}apwell milling factory},
  in: \bibinfo{booktitle}{2022 IEEE PES/IAS PowerAfrica}, pp.
  \bibinfo{pages}{1--5}.
\bibitem[{Vilalta et~al.(2019)Vilalta, Wan and
  Patnaik}]{Centrifugal_Vilalta2019}
\bibinfo{author}{Vilalta, P.C.}, \bibinfo{author}{Wan, H.},
  \bibinfo{author}{Patnaik, S.S.}, \bibinfo{year}{2019}.
\newblock \bibinfo{title}{Centrifugal compressor performance prediction using
  {G}aussian process regression and artificial neural networks}, in:
  \bibinfo{booktitle}{Volume 8: Heat Transfer and Thermal Engineering},
  \bibinfo{publisher}{American Society of Mechanical Engineers}.
\bibitem[{Wu et~al.(2000)Wu, Rios-Mercado, Boyd and Scott}]{Model_Wu2000}
\bibinfo{author}{Wu, S.}, \bibinfo{author}{Rios-Mercado, R.Z.},
  \bibinfo{author}{Boyd, E.A.}, \bibinfo{author}{Scott, L.R.},
  \bibinfo{year}{2000}.
\newblock \bibinfo{title}{Model relaxations for the fuel cost minimization of
  steady-state gas pipeline networks}.
\newblock \bibinfo{journal}{Mathematical and Computer Modelling}
  \bibinfo{volume}{31}, \bibinfo{pages}{197--220}.
\bibitem[{Xenos(2015)}]{Optimal_Xenos2015}
\bibinfo{author}{Xenos, D.P.}, \bibinfo{year}{2015}.
\newblock \bibinfo{title}{Optimal Operation of Industrial Compressor Stations
  in Systems with Large Energy Consumption}.
\newblock Ph.D. thesis.
\newblock \bibinfo{note}{Imperial College London}.
\bibitem[{Zagorowska and Thornhill(2020)}]{Influence_Zagorowska2019}
\bibinfo{author}{Zagorowska, M.}, \bibinfo{author}{Thornhill, N.F.},
  \bibinfo{year}{2020}.
\newblock \bibinfo{title}{Influence of compressor degradation on optimal
  load-sharing}.
\newblock \bibinfo{journal}{Computers and Chemical Engineering}
  \bibinfo{volume}{143}, \bibinfo{pages}{107104}.

\end{thebibliography}
  
  \end{document}